\documentclass[aps,prx,twocolumn,footinbib,showkeys,nobalancelastpage]{revtex4-2}

\usepackage{times}
\usepackage{latexsym}
\usepackage{graphicx}
\usepackage{amsmath,amssymb}
\usepackage{tensor}
\usepackage{verbatim,bbm}
\usepackage{textcomp}
\usepackage{indentfirst}
\usepackage{physics}
\usepackage{braket}
\usepackage{float}
\usepackage{upgreek}
\usepackage{epstopdf}
\usepackage{CJK}
\usepackage{esint}
\usepackage{color}
\usepackage{subfigure}
\usepackage{amsfonts}
\usepackage{footmisc}
\usepackage{scrextend}
\usepackage{multirow}
\usepackage[hyperfootnotes=false]{hyperref}
\usepackage[english]{babel}
\usepackage{url}
\usepackage[T1]{}

\begin{document}

\title{Bounds on amplitude damping channel discrimination}
\author{Jason L. Pereira}
\author{Stefano Pirandola}
\affiliation{Department of Computer Science, University of York, York YO10 5GH, UK}
\date{\today}

\begin{abstract}
Amplitude damping (AD) channels are good models for many physical scenarios, and so the development of protocols to discriminate between them is an important task in quantum information science. It is therefore important to bound the performance of such protocols. Since adaptivity has been shown to improve the performance of discrimination protocols, bounds on the distinguishability of AD channels must take this into account. In this paper, we use both channel simulation and a bound based on the diamond norm to significantly tighten the upper bound on the trace norm between the possible outputs of binary channel discrimination protocols acting on AD channels (and hence the lower bound on the error probability of such protocols). The diamond norm between any two AD channels is found analytically, giving the optimal error probability for a one-shot discrimination protocol. We also present a tighter lower bound on the achievable trace norm between protocol outputs (and a corresponding upper bound on the achievable error probability). The upper and lower bounds are compared with existing bounds and then applied to quantum hacking and biological quantum sensing scenarios.
\end{abstract}

\maketitle

\section{Introduction}
Quantum channel discrimination is the task of determining which quantum channel is present out of a set of possible channels. It is an important task in quantum information because it has many physical applications~\cite{pirandola_advances_2018}. An example in quantum communications is quantum hacking~\cite{scarani_black_2014,jain_attacks_2016,pirandola_advances_2020}, where Eve may wish to determine aspects of the settings of Alice's and Bob's devices by probing them via side-channels. If the settings affect the quantum channel that the probes would pass through, Eve could carry out a discrimination protocol between the possible channels, and therefore the possible settings.

Within quantum channel discrimination, one well-studied task is binary discrimination. This is discrimination between only two possible channels. For equal prior probabilities, the error probability in distinguishing between the two possible output states of a discrimination protocol is known exactly, in terms of the trace norm (the Helstrom bound)~\cite{helstrom_quantum_1969}. One application of binary discrimination is in quantum illumination~\cite{lloyd_enhanced_2008,tan_quantum_2008,shapiro_quantum_2009,barzanjeh_microwave_2015,Zhuang17,Zhuang17b,Wilde17,QuntaoLIDARS,Depalma18,NairGu,Athena2020,Lopaeva13,ZhangQI13,ZhangQI15}, where a device must discriminate between the presence and the absence of an object. Another application is the protocol of quantum reading~\cite{qreading} (see the recent review~\cite{pirandola_advances_2018} and references therein).

Adaptive protocols, where subsequent probes can be dependent on measurements carried out on previous probes, have proven to be more powerful than non-adaptive protocols for certain pairs of channels~\cite{harrow_adaptive_2010}. Note that this is not the case for all channels: adaptivity has been shown to be unnecessary for discriminating between unitary channels~\cite{acin_statistical_2001} and to not help the asymptotic error rate for discriminating between classical channels~\cite{hayashi_discrimination_2009}. Nonetheless, for pairs of channels that have not been proven to not require adaptivity for optimal discrimination, adaptivity must be taken into account. This has necessitated the study of the most general adaptive protocols, in order to establish ultimate bounds on the minimum achievable error probability for quantum channel discrimination~\cite{pirandola_advances_2018}. Quantum channel simulation is a powerful tool for establishing these ultimate bounds~\cite{pirandola_fundamental_2017,pirandola_fundamental_2019,cope_simulation_2017,pirandola_conditional_2019,convex}.

Pure loss channels constitute an important class of quantum channels. They can be used as models in many situations in which the environmental noise is low. Examples include quantum communications~\cite{pirandola_fundamental_2017} and quantum metrology~\cite{Sam94} (where the parameter being measured could be the loss of the channel). Amplitude damping (AD) channels are qubit channels that act similarly to lossy channels: they can be regarded as lossy channels that only act on qubit states. In physics, they are good models for energy dissipation in qubit systems~\cite{nielsen_quantum_2011}, and in quantum information, they can model low noise scenarios where the number of photons passing through a quantum channel is also low. They have also been used as a model for the transfer of a qubit through a spin chain~\cite{bose_quantum_2003}.

AD channel discrimination was briefly discussed in Ref.~\cite{pirandola_fundamental_2019}. The authors presented a lower bound on the discrimination error of adaptive protocols based on channel simulation using port-based teleportation (PBT) and an upper bound based on (non-adaptive) probing with Bell states. Ref.~\cite{zhuang_ultimate_2020} investigated the multiple channel discrimination case (i.e. discrimination between more than two possible channels) of channel position finding over a sequence of AD channels. Ref.~\cite{katariya_evaluating_2020} investigated adaptive discrimination of generalised amplitude damping channels (which can be reduced to AD channel discrimination by setting the noise parameter to $0$) via numerical minimisation methods. Ref.~\cite{rexiti_discriminating_2021} presented results relating to feedback-assisted measurements for non-adaptive AD channel discrimination.

In this work, we find various upper bounds on the trace norm between the possible output states of any adaptive discrimination protocol that is discriminating between AD channels. We therefore lower bound the optimal error probabilities for binary discrimination tasks between AD channels. We present two new upper bounds based on PBT channel simulation, with different resource states from those used in Ref.~\cite{pirandola_fundamental_2019}, and an upper bound based on the diamond norm. The latter bound is also applicable to a wider variety of discrimination tasks. We give an analytical expression for the exact diamond norm between any pair of AD channels. We present a tighter lower bound on the optimal trace norm for discrimination protocols between AD channels, which is well approximated by an approximate bound based on the quantum Cram\'{e}r-Rao bound (QCRB), given in Ref.~\cite{spedalieri_detecting_2020}, for a large number of channel uses.

We numerically investigate the presented bounds, and compare them with the existing bounds found in Ref.~\cite{pirandola_fundamental_2019}. We then apply the bounds to a quantum hacking scenario, in which Eve is attempting to learn Alice's bases for BB84, using a side-channel. We also apply the bounds to a biological quantum metrology scenario, in which a sample must be probed with low energy quantum states, in order to detect the presence of bacteria or to discriminate between two types of bacteria.

In Section~\ref{section: analytical}, we describe the task of binary AD channel discrimination via an adaptive protocol and present the various bounds. We also calculate the diamond norm between AD channels. In Section~\ref{section: numerical}, we compare the bounds with existing bounds and apply them to two different scenarios. Finally, in Section~\ref{section: conclusion}, we present our conclusions.

\section{Analytical results}\label{section: analytical}
Suppose we are given an AD channel, $\mathcal{C}$, which we know to have a transmissivity, $\eta$, equal to either $\eta_X$ or $\eta_Y$, and wish to determine which of these two values $\eta$ takes. AD channels are qubit channels and are defined by having the Kraus operators
\begin{align}
&K_0=\left|0\right>\left<0\right|+\sqrt{\eta}\left|1\right>\left<1\right|,\\
&K_1=\sqrt{1-\eta}\left|0\right>\left<1\right|.
\end{align}
Equivalently, they can be described by the Choi matrix
\begin{equation}
C_{\mathrm{AD}}=\begin{pmatrix}
\frac{1}{2} &0 &0 &\frac{\sqrt{\eta}}{2}\\
0 &0 &0 &0\\
0 &0 &\frac{1-\eta}{2} &0\\
\frac{\sqrt{\eta}}{2} &0 &0 &\frac{\eta}{2}
\end{pmatrix}.
\end{equation}
The Choi matrix of a qubit channel is the state obtained by sending one half of a Bell pair through the channel. In this case, the Bell pair being used is $\frac{1}{\sqrt{2}}(\left|00\right>+\left|11\right>)$; in the context of Choi resources for PBT (as discussed in detail in Subsection~\ref{subsection: PBT upper bounds}), the Bell pair used to generate Choi matrices is $\frac{1}{\sqrt{2}}(\left|01\right>-\left|10\right>)$.

Note that an AD channel is the qubit version of a pure loss channel, in that the pointwise application of a hard energy constraint of one photon and a pure loss channel with transmissivity $\eta$ reduces to an AD channel with transmissivity $\eta$ (or damping probability $1-\eta$). Suppose we are allowed to carry out any protocol involving our channel, but with a maximum of $N$ channel uses. Let $\mathcal{C}_X$ be the AD channel with a transmissivity of $\eta_X$ and let $\mathcal{C}_Y$ be the AD channel with a transmissivity of $\eta_Y$. Our task is to carry out the optimal protocol for discriminating between $\mathcal{C}_X$ and $\mathcal{C}_Y$, subject to the constraint on the total number of channel uses.

A general protocol consists of quantum operations on some initial state, followed by a channel use, followed by further operations (which can include measurements) and further channel uses, until a total of $N$ channel uses have occurred~\cite{pirandola_fundamental_2019}. Note that the initial state can be arbitrarily large and may include idler modes. At this point, a final set of quantum operations is carried out, and then a measurement is made on the final state, which we will label as $\rho^{N,out}_i$ in the case in which the channel is $\mathcal{C}_i$. Note that this protocol is allowed to be adaptive, meaning that each step in the protocol can depend on previous steps. We define the optimal protocol as the protocol for which we maximise the trace norm between $\rho^{N,\mathrm{out}}_X$ and $\rho^{N,\mathrm{out}}_Y$ and then carry out the most discriminating measurement possible. This optimal value of the trace norm is denoted by $D^{\mathrm{opt},N}_{\mathcal{C}_X \mathcal{C}_Y}$. If we have $\mathcal{C}_X$ and $\mathcal{C}_Y$ with equal probabilities, this is the protocol that minimises the probability of error in identifying which channel we have. It is also worth noting that, for the optimal protocol, we can assume without loss of generality that all of the operations between channel uses are unitaries, as any other operations (such as quantum channels, of which measurements are a special case) can be modelled as unitaries by allowing the user of the protocol to hold the distillation of all operations performed. This cannot decrease the trace norm between output states.

\subsection{Bounding the maximum trace norm using channel simulation}\label{subsection: bounding with channel simulation}
We now apply the technique of channel simulation~\cite{pirandola_fundamental_2017,convex,pirandola_fundamental_2019}. Suppose we have a qubit quantum processor $\mathcal{Q}(\pi)$, which takes the resource state $\pi$ as a program and enacts the channel $\mathcal{C}_{\mathcal{Q}(\pi)}$ on an input qubit, via some set of trace-preserving quantum operations. Suppose also that there exist program states $\pi_X$ and $\pi_Y$, such that the enacted channels, $\mathcal{C}_{\mathcal{Q}(\pi_X)}$ and $\mathcal{C}_{\mathcal{Q}(\pi_Y)}$, are sufficiently close to the two AD channels that we want to discriminate between. More precisely, suppose we can write
\begin{align}
&\left\|\mathcal{C}_{\mathcal{Q}(\pi_X)}-\mathcal{C}_X\right\|_{\diamond}\leq \epsilon_X,\\
&\left\|\mathcal{C}_{\mathcal{Q}(\pi_Y)}-\mathcal{C}_Y\right\|_{\diamond}\leq \epsilon_Y,
\end{align}
where we have used the diamond norm between the channels. This is the maximum of the trace norm between the outputs of the channels, maximised over all input states (including those with idlers). Then, we replace the $N$ channel uses in our discrimination protocol with the channel enacted by the processor (with program state $\pi_{X(Y)}$ in the case in which the channel is $\mathcal{C}_{X(Y)}$), and call the output state of the resulting protocol $\rho^{N,\mathrm{out}}_{\mathcal{Q}(\pi_{X(Y)})}$. By using the contractivity of the trace norm under quantum channels and the triangle inequality (see Subsection \ref{subsection: trivial bound} for a more detailed derivation along similar lines), we can then write
\begin{align}
&\left\|\rho^{N,\mathrm{out}}_{\mathcal{Q}(\pi_X)}-\rho^{N,\mathrm{out}}_{X}\right\|_{1}\leq N\epsilon_X,\\
&\left\|\rho^{N,\mathrm{out}}_{\mathcal{Q}(\pi_Y)}-\rho^{N,\mathrm{out}}_{Y}\right\|_{1}\leq N\epsilon_Y.
\end{align}

Using the fact that all of the operations are trace-preserving, and the only difference between the two cases is the initial program state, we can write
\begin{align}
\left\|\rho^{N,\mathrm{out}}_{\mathcal{Q}(\pi_X)}-\rho^{N,\mathrm{out}}_{\mathcal{Q}(\pi_Y)}\right\|_{1}&\leq \left\|\pi_X^{\otimes N}-\pi_Y^{\otimes N}\right\|_{1}\\
&\leq 2\sqrt{1-F\left(\pi_X^{\otimes N},\pi_Y^{\otimes N}\right)^2},\label{eq: fuchs van der graaf}
\end{align}
where $F(\rho_1,\rho_2)$ is the quantum fidelity, defined by
\begin{align}
F(\rho_1,\rho_2)=\mathrm{Tr}\sqrt{\sqrt{\rho_1}\rho_2\sqrt{\rho_1}}.
\end{align}

When bounding the trace norm between $N$ copies of each program state, one might also consider using the quantum Chernoff bound (QCB)~\cite{audenaert_discriminating_2007}. The QCB in the non-logarithmic form is defined by
\begin{equation}
Q(\rho_1,\rho_2)=\min_{0\leq s\leq 1} \mathrm{Tr}\left[\rho_1^s \rho_2^{1-s}\right].
\end{equation}
We can then bound the trace norm between two states, $D$, with
\begin{equation}
2(1-F)\leq 2(1-Q)\leq D \leq 2\sqrt{1-F^2} \leq 2\sqrt{1-Q^2},
\end{equation}
since $Q(\rho_1,\rho_2)\leq F(\rho_1,\rho_2)$~\cite{calsamiglia_quantum_2008}. Since we are interested in upper bounding the trace norm between two program states, the Fuchs-van der Graaf relation (in Eq.~(\ref{eq: fuchs van der graaf})) gives a tighter bound (although the QCB gives a tighter lower bound). Note that the QCB approaches the trace distance asymptotically (in terms of error exponent), but it is generally non-tight for a finite number of channel uses.

Using the multiplicativity of the fidelity with respect to tensor products, we get
\begin{align}
\left\|\rho^{N,\mathrm{out}}_{\mathcal{Q}(\pi_X)}-\rho^{N,\mathrm{out}}_{\mathcal{Q}(\pi_Y)}\right\|_{1}\leq 2\sqrt{1-F\left(\pi_X,\pi_Y\right)^{2N}}.
\end{align}
Finally, using the triangle inequality, we write
\begin{align}
&D^{\mathrm{opt},N}_{\mathcal{C}_X \mathcal{C}_Y}\leq N\epsilon_{XY}+2\sqrt{1-F\left(\pi_X,\pi_Y\right)^{2N}},\\
&\epsilon_{XY}=\epsilon_X+\epsilon_Y,\label{eq: trace norm bound}
\end{align}
where $D^{\mathrm{opt},N}_{\mathcal{C}_X \mathcal{C}_Y}$ is maximised over all possible protocols.

The trace norm between two states is related to the maximum probability of successfully discriminating between them, $p^{\mathrm{succ}}$ via
\begin{align}
p^{\mathrm{succ}}=\frac{1}{2}+\frac{D}{4},
\end{align}
and so we have an upper bound on the probability of discriminating between two amplitude damping channels $\mathcal{C}_X$ and $\mathcal{C}_Y$, which holds over all possible adaptive protocols. Alternatively, by defining
\begin{align}
p^{\mathrm{err}}&=1-p^{\mathrm{succ}}\\
&=\frac{1}{2}-\frac{D}{4},\label{eq: error probability}
\end{align}
we have a lower bound on the error probability.

Note that the tightness of this bound depends both on the chosen program states, $\pi_X$ and $\pi_Y$, and on the quantum processor, $\mathcal{Q}$, used. In order to attain a tight bound, we need to both minimise the simulation errors, $\epsilon_X$ and $\epsilon_Y$, and minimise the trace norm between the program states simultaneously. For instance, we could conceive of a trivial quantum processor that measures the program state in the computational basis and then, depending on the outcome of the measurement, enacts either $\mathcal{C}_X$ or $\mathcal{C}_Y$. Choosing the program states $\left|0\right>$ and $\left|1\right>$, we get $\epsilon_{X(Y)}=0$, but the trace norm between the program states is maximised, and hence our bound is too large. More useful bounds can be found with processors that use PBT, as discussed in Subsection~\ref{subsection: processors}, and with a different trivial processor, as discussed in Subsection~\ref{subsection: trivial bound}.

\subsection{Quantum processors for AD channel simulation}\label{subsection: processors}
As previously mentioned, the tightness of the bound depends on the quantum processor and program states used to simulate the channels. We wish to minimise the simulation error whilst keeping our program states as similar to each other as possible, in order to achieve the tightest possible bound.

One idea that may be intuitively appealing is to use (standard) quantum teleportation~\cite{bennett_teleporting_1993} to simulate the AD channels. For certain qubit channels (namely, Pauli channels), quantum teleportation using the Choi matrix of the channel as a resource (program state) can perfectly simulate the channel (with a simulation error of 0).

The issue with this is that standard quantum teleportation cannot simulate non-Pauli channels~\cite{bowen_teleportation_2001}, and so we would have a very high simulation error. This would result in a bound that would be too loose to be useful.

One alternative is to use port-based teleportation~\cite{ishizaka_asymptotic_2008,ishizaka_quantum_2009}. PBT uses a combined measurement (the square-root measurement) on an input state and $m$ ports, held by the sender, to teleport the input state to one of $m$ ports, held by the receiver. The receiver then traces over the remaining ports. The process is discussed in more detail in Refs.~\cite{ishizaka_asymptotic_2008,ishizaka_quantum_2009,pereira_characterising_2021}. The program state is the shared resource state of $2m$ qubits, $m$ of which constitute the sender's ports and $m$ of which constitute the receiver's ports.

A possible program state, in this case, is $m$ copies of the Choi matrix of the AD channel. It is known that in the asymptotic limit of $m\to\infty$, such a simulation becomes perfect. The issue with this is that the trace norm between the program states of the two possible channels increases as the number of copies increases, and so we cannot take the asymptotic limit of $m$. Instead, we can accept some small but non-zero simulation error, and try to find the optimal value of $m$ to minimise the total value of the bound.

This is the approach taken by Ref.~\cite{pirandola_fundamental_2019}, for calculating a lower bound for the error probability of discriminating between two AD channels (i.e. the same type of bound that we want to calculate here). We will call the family of bounds that come from PBT simulations using the Choi matrix of the simulated channels as a resource the standard Choi bounds (and will implicitly assume that the optimal value of $m$ has been chosen).

In fact, for finite $m$, there are program states that simulate AD channels better than $m$ copies of the Choi matrix of the simulated channel. This was discussed in Ref.~\cite{pereira_characterising_2021}, in which two classes of resource states capable of providing better simulations of AD channels were presented.

The first class uses $m$ copies of the Choi matrix of a different AD channel from the one being simulated as a resource. Specifically, to simulate an AD channel with transmissivity $\eta$, we use $m$ copies of the Choi matrix of the AD channel with transmissivity $\eta'$, where
\begin{align}
\eta'=\frac{\eta}{1-\xi_m}.\label{eq: improved Choi eta}
\end{align}
$\xi_m$ is the PBT coefficient for $m$ ports, as defined in Eq.~(11) of Ref.~\cite{pirandola_fundamental_2019}, and represents the depolarisation probability when carrying out PBT with a maximally entangled resource state. As such, it is a number between 0 and 1, and consequently $\eta'>\eta$. Our notation here differs from Ref.~\cite{pereira_characterising_2021}, since we are characterising the AD channels with $\eta$ rather than the damping probability (which is $1-\eta$). Note that we also require $\eta\leq 1-\xi_m$. We will call the bounds deriving from PBT using this resource state the improved Choi bounds.

The second class uses pure resource states, parametrised by a parameter $a$, that take the form
\begin{align}
R^{alt}(a)=\left(\sqrt{a}\left|01\right> - \sqrt{1-a}\left|10\right>\right)^{\otimes m}.\label{eq: alt resource}
\end{align}
An advantage that comes from the fact that this resource state is pure is that the trace distance between different program states is analytically calculable (since the upper bound coming from the fidelity is tight). The value of the parameter $a$ is determined by both the damping probability of the AD channel that is being simulated and by the number of ports, $m$, and is chosen so as to minimise the simulation error. We will call the bounds deriving from PBT using this resource state the alternative resource bounds.

The upper bounds based on program states of these types are calculated in Subsection~\ref{subsection: PBT upper bounds}. In all three cases, we must tune $m$ so as to obtain the tightest bound possible.

\subsection{The trivial bound}\label{subsection: trivial bound}
We can also formulate a bound based on a trivial processor that simply always enacts the channel $\mathcal{C}_X$. In this case, we have
\begin{align}
&\epsilon_X=0,\\
&\epsilon_Y=\left\|\mathcal{C}_X-\mathcal{C}_Y\right\|_{\diamond},\\
&\left\|\pi_X^{\otimes M}-\pi_Y^{\otimes M}\right\|_{1}=0.
\end{align}
In other words, the bound in Eq.~(\ref{eq: trace norm bound}) simply becomes
\begin{align}
&D^{\mathrm{opt},N}_{\mathcal{C}_X \mathcal{C}_Y}\leq N D^{\diamond}_{\mathcal{C}_X\mathcal{C}_Y},\label{eq: trivial bound}\\
&D^{\diamond}_{\mathcal{C}_X\mathcal{C}_Y}=\left\|\mathcal{C}_X-\mathcal{C}_Y\right\|_{\diamond}.
\end{align}
This is $N$ times the diamond distance between the two channels that we are trying to distinguish between. Note that this bound is not specific to AD channels, and could be applied to any binary discrimination task.

In fact, we can write an alternative and simpler proof that this bound holds. Let
\begin{align}
S(N,m)=\{\mathcal{C}_X,\mathcal{C}_X,...\mathcal{C}_X,\mathcal{C}_Y,\mathcal{C}_Y,...\mathcal{C}_Y\}
\end{align}
be a sequence of $N$ channels that are either $\mathcal{C}_X$ or $\mathcal{C}_Y$. Specifically, the first $m$ channels are $\mathcal{C}_X$ and the next $N-m$ channels are $\mathcal{C}_Y$. Then let $\mathcal{P}(S(N,m))$ be the output of a fully general and potentially adaptive protocol $\mathcal{P}$, which has a total of $N$ channel uses, where the $i$-th channel use involves sending the signal through the channel that is the $i$-th element of $S$. E.g. if $S(3,2)=\mathcal{C}_X,\mathcal{C}_X,\mathcal{C}_Y$, $\mathcal{P}(S)$ is the output of a discrimination protocol when the channel that we are trying to identify as either $\mathcal{C}_X$ or $\mathcal{C}_Y$ (and which the protocol assumes is always the same) is $\mathcal{C}_X$ for the first two channel uses and is $\mathcal{C}_Y$ for the final channel use. We then have
\begin{align}
&\mathcal{P}(S(N,N))=\mathcal{P}({\mathcal{C}_X,\mathcal{C}_X,...\mathcal{C}_X})=\rho^{N,\mathrm{out}}_X,\\
&\mathcal{P}(S(N,0))=\mathcal{P}({\mathcal{C}_Y,\mathcal{C}_Y,...\mathcal{C}_Y})=\rho^{N,\mathrm{out}}_Y.
\end{align}

We therefore want to upper bound the trace distance between $\mathcal{P}(S(N,N))$ and $\mathcal{P}(S(N,0))$. We start by writing
\begin{align}
\left\|\mathcal{P}(S(N,N))-\mathcal{P}(S(N,N-1))\right\|_{1}\leq D^{\diamond}_{\mathcal{C}_X\mathcal{C}_Y}.
\end{align}
This is due to the fact that the states are identical prior to the final channel use, the states immediately after the final channel use cannot be further apart than the diamond distance between the two channels and any subsequent post-processing is the same in both cases, and so cannot increase the trace distance between the states.

By a similar argument we have
\begin{align}
\left\|\mathcal{P}(S(N,N-1))-\mathcal{P}(S(N,N-2))\right\|_{1}\leq D^{\diamond}_{\mathcal{C}_X\mathcal{C}_Y},
\end{align}
and generalising, we can write
\begin{align}
\left\|\mathcal{P}(S(N,i))-\mathcal{P}(S(N,i-1))\right\|_{1}\leq D^{\diamond}_{\mathcal{C}_X\mathcal{C}_Y}.
\end{align}
Then, using the triangle inequality, we can write
\begin{align}
\left\|\mathcal{P}(S(N,i))-\mathcal{P}(S(N,i-j))\right\|_{1}\leq (i-j)D^{\diamond}_{\mathcal{C}_X\mathcal{C}_Y},
\end{align}
and therefore
\begin{align}
\left\|\mathcal{P}(S(N,N))-\mathcal{P}(S(N,0))\right\|_{1}\leq N D^{\diamond}_{\mathcal{C}_X\mathcal{C}_Y},
\end{align}
as required.

The diamond norm between any two AD channels is presented in Subsection~\ref{subsection: diamond norm}.

\subsection{Calculating the diamond norm between two AD channels}\label{subsection: diamond norm}
We start by making an ansatz that the exact diamond norm between two AD channels can be achieved using a state of the form
\begin{align}
\left|\phi(t)\right>=\sqrt{t}\left|00\right>+\sqrt{1-t}\left|11\right>,\label{eq: probe gen form}
\end{align}
where $0\leq t\leq1$. The trace norm between two AD channels for such a state, $D^{\left|\phi(t)\right>,1}_{\mathcal{C}_X \mathcal{C}_Y}$, is then given by
\begin{align}
&D^{\left|\phi(t)\right>,1}_{\mathcal{C}_X \mathcal{C}_Y}=|\eta_X-\eta_Y|(1-t)\left(1+\sqrt{1+\frac{4t}{(1-t)x^2}}\right),\\
&x=\sqrt{\eta_X}+\sqrt{\eta_Y}.
\end{align}

Defining $t_{\mathrm{max}}$ as the value of $t$ that maximises $D^{\left|\phi(t)\right>,1}_{\mathcal{C}_X \mathcal{C}_Y}$, we find
\begin{align}
t_{\mathrm{max}}=\mathrm{max}\{0,1-\frac{1}{2-x}\}.
\end{align}
From this, we can see that the problem is split into two regimes: one in which $t_{\mathrm{max}}=0$ and one in which $t_{\mathrm{max}}>0$. Note that for the extremal case of $t=0$, the idler mode is not necessary, since the state given by Eq.~(\ref{eq: probe gen form}) is separable for this parameter value. Consequently, probing with $\left|1\right>$ is equivalent to probing with $\left|11\right>$. The first regime occurs when
\begin{align}
\sqrt{\eta_x}+\sqrt{\eta_Y}>1.
\end{align}
We can then calculate the trace norms for each regime:
\begin{align}
&D^{\diamond,t=0}_{\mathcal{C}_X\mathcal{C}_Y}=2|\eta_X-\eta_Y|,\label{eq: diamond norm t=0}\\
&D^{\diamond,t>0}_{\mathcal{C}_X\mathcal{C}_Y}=\frac{2|\sqrt{\eta_X}-\sqrt{\eta_Y}|}{2-(\sqrt{\eta_X}+\sqrt{\eta_Y})}.\label{eq: diamond norm t>0}
\end{align}

The next step is to prove that the expressions in Eqs.~(\ref{eq: diamond norm t=0}) and~(\ref{eq: diamond norm t>0}) are the diamond norms in each regime. We do this using semidefinite programming. Ref.~\cite{watrous_simpler_2013} showed that finding the diamond norm can be reduced to a semidefinite programming problem. In a semidefinite programming problem, some matrices must be chosen, subject to constraints, to maximise or minimise a quantity that is dependent on these matrices. More specifically, every problem consists of a primal and a dual problem. Each valid solution to the primal problem provides a lower bound to the quantity, and so one maximises over the primal problem. Each valid solution to the dual problem provides an upper bound to the quantity, and so one minimises over it. Therefore, in order to show that Eqs.~(\ref{eq: diamond norm t=0}) and~(\ref{eq: diamond norm t>0}) give the diamond norm, we must find matrices satisfying the constraints of the dual problem for the diamond norm that give the expressions in Eqs.~(\ref{eq: diamond norm t=0}) and~(\ref{eq: diamond norm t>0}) as the diamond norm. The dual problem is to find positive matrices $Y_0$ and $Y_1$ that satisfy the constraint
\begin{align}
M=\begin{pmatrix}
Y_0 &-J(\mathcal{C}_X,\mathcal{C}_Y)\\
-J(\mathcal{C}_X,\mathcal{C}_Y) &Y_1
\end{pmatrix}>0,
\end{align}
where $J$ is the Choi matrix of channel $\mathcal{C}_X$ minus the Choi matrix of channel $\mathcal{C}_Y$, multiplied by the dimension of the input system (which, in our case, is $2$). The upper bound on the diamond norm, which we must then minimise, is
\begin{align}
D^{\diamond}_{\mathcal{C}_X\mathcal{C}_Y}\leq \frac{\left\|\mathrm{Tr}_{S}(Y_0)\right\|_{\infty}+\left\|\mathrm{Tr}_{S}(Y_1)\right\|_{\infty}}{2},
\end{align}
where the partial trace is taken over the signal (rather than the idler) mode, and the norm is the operator norm (i.e. the largest eigenvalue). In our case, we have
\begin{align}
J(\mathcal{C}_X,\mathcal{C}_Y)=\begin{pmatrix}
0 &0 &0 &\sqrt{\eta_X}-\sqrt{\eta_Y}\\
0 &0 &0 &0\\
0 &0 &\eta_Y-\eta_X &0\\
\sqrt{\eta_X}-\sqrt{\eta_Y} &0 &0 &\eta_X-\eta_Y
\end{pmatrix}.
\end{align}

Let us first consider the $t=0$ case. Consider the matrices
\begin{align}
&Y_0^{t=0}=\begin{pmatrix}
D^{\diamond,t=0}_{\mathcal{C}_X\mathcal{C}_Y} &0 &0 &|\sqrt{\eta_X}-\sqrt{\eta_Y}|\\
0 &0 &0 &0\\
0 &0 &\frac{1}{2}D^{\diamond,t=0}_{\mathcal{C}_X\mathcal{C}_Y} &0\\
|\sqrt{\eta_X}-\sqrt{\eta_Y}| &0 &0 &\frac{1}{2}D^{\diamond,t=0}_{\mathcal{C}_X\mathcal{C}_Y}
\end{pmatrix},\\
&Y_1^{t=0}=Y_0^{t=0}.
\end{align}
We can immediately see that
\begin{align}
\mathrm{Tr}_S\left(Y_0^{t=0}\right)=\mathrm{Tr}_S\left(Y_1^{t=0}\right)=D^{\diamond,t=0}_{\mathcal{C}_X\mathcal{C}_Y}\begin{pmatrix}
1 &0\\
0 &1\\
\end{pmatrix},
\end{align}
and so the upper bound on the diamond norm coming from this solution is equal to the expression in Eq.~(\ref{eq: diamond norm t=0}). The distinct, non-zero eigenvalues of $M^{t=0}$ are
\begin{align}
&e_{M^{t=0}}^1=2|\eta_X-\eta_Y|,\\
&e_{M^{t=0}}^2=2\left(|\eta_X-\eta_Y|-\left|\sqrt{\eta_X}-\sqrt{\eta_Y}\right|\right),\\
&e_{M^{t=0}}^3=2\left(|\eta_X-\eta_Y|+\left|\sqrt{\eta_X}-\sqrt{\eta_Y}\right|\right),
\end{align}
the smallest of which is $e_{M^{t=0}}^2$. Since $e_{M^{t=0}}^2>0$ for the regime in which $t=0$ (i.e. for $x>1$), $M^{t=0}>0$ in this regime, as required. The non-zero eigenvalues of $Y_0^{t=0}$ (and $Y_1^{t=0}$) are
\begin{align}
&e_{Y^{t=0}}^1=|\eta_X-\eta_Y|,\\
&e_{Y^{t=0}}^2=\frac{|\eta_X-\eta_Y|}{2}\left(3-\sqrt{1+\frac{4}{x^2}}\right),\\
&e_{Y^{t=0}}^3=\frac{|\eta_X-\eta_Y|}{2}\left(3+\sqrt{1+\frac{4}{x^2}}\right).
\end{align}
$e_{Y^{t=0}}^2$ is the smallest of these, and $e_{M^{t=0}}^2>0$ in the regime in which $t=0$, so both $Y_0^{t=0}$ and $Y_1^{t=0}$ are positive. Therefore, Eq.~(\ref{eq: diamond norm t=0}) gives the exact diamond norm for the $t=0$ regime.

Next, we consider the $t>0$ case. Consider the matrices
\begin{align}
&Y_0^{t>0}=\begin{pmatrix}
D^{\diamond,t>0}_{\mathcal{C}_X\mathcal{C}_Y} &0 &0 &\frac{|\eta_X-\eta_Y|}{2-x}\\
0 &0 &0 &0\\
0 &0 &\frac{1}{2}D^{\diamond,t=0}_{\mathcal{C}_X\mathcal{C}_Y} &0\\
\frac{|\eta_X-\eta_Y|}{2-x} &0 &0 &D^{\diamond,t>0}_{\mathcal{C}_X\mathcal{C}_Y}-\frac{1}{2}D^{\diamond,t=0}_{\mathcal{C}_X\mathcal{C}_Y}
\end{pmatrix},\\
&Y_1^{t>0}=Y_0^{t>0}.
\end{align}
Tracing over the signal mode, we get
\begin{align}
\mathrm{Tr}_S\left(Y_0^{t>0}\right)=\mathrm{Tr}_S\left(Y_1^{t>0}\right)=D^{\diamond,t>0}_{\mathcal{C}_X\mathcal{C}_Y}\begin{pmatrix}
1 &0\\
0 &1\\
\end{pmatrix},
\end{align}
and so the upper bound on the diamond norm coming from this solution is equal to the expression in Eq.~(\ref{eq: diamond norm t>0}). The non-zero eigenvalues of $M^{t>0}$ are
\begin{align}
&e_{M^{t=0}}^1=2|\eta_X-\eta_Y|,\\
&e_{M^{t=0}}^2=\frac{4|\sqrt{\eta_X}-\sqrt{\eta_Y}|}{2-x},\\
&e_{M^{t=0}}^3=\frac{4|\sqrt{\eta_X}-\sqrt{\eta_Y}|}{2-x}-2|\eta_X-\eta_Y|,
\end{align}
which are all positive. The non-zero eigenvalues of $Y_0^{t>0}$ (and $Y_1^{t>0}$) are
\begin{align}
&e_{Y^{t>0}}^1=|\eta_X-\eta_Y|,\\
\begin{split}
&e_{Y^{t>0}}^2=\frac{|\sqrt{\eta_X}-\sqrt{\eta_Y}|}{2(2-x)}\Big(3+(1-x)^2\\
&\; \; \; \; \; \; \; \; \; \; \; \; \; \; \; \; \; \; \; \; \; \; \; \; \; \; \; \; \; \; \; \; \; \; \; \; \; \; \;
-x\sqrt{4+(2-x)^2}\Big),
\end{split}\\
\begin{split}
&e_{Y^{t>0}}^3=\frac{|\sqrt{\eta_X}-\sqrt{\eta_Y}|}{2(2-x)}\Big(3+(1-x)^2\\
&\; \; \; \; \; \; \; \; \; \; \; \; \; \; \; \; \; \; \; \; \; \; \; \; \; \; \; \; \; \; \; \; \; \; \; \; \; \; \;
+x\sqrt{4+(2-x)^2}\Big).
\end{split}
\end{align}
These are again all positive, proving that Eq.~(\ref{eq: diamond norm t>0}) gives the exact diamond norm for the $t>0$ regime.

A logical next step would be to calculate the diamond norm between multiple uses of two AD channels, i.e. between the two channels $\mathcal{C}_X^{\otimes N}$ and $\mathcal{C}_Y^{\otimes N}$. However, this is a more difficult task. Numerically, we find that input states of the form $\left|1\right>^{\otimes N}$ achieve the diamond norm in some cases, as for the single use case, but that the regimes are more complicated to characterise. Further, for large numbers of channel uses, numerically finding the diamond norm via semidefinite programming is computationally expensive.

\subsection{Lower bounds on the optimal trace norm}\label{subsection: trace norm lower bounds}
It is helpful to also find lower bounds on the maximum trace distance between protocol outputs, $D^{\mathrm{opt},N}_{\mathcal{C}_X \mathcal{C}_Y}$ since this allows us to assess how tight our upper bounds are. One option is to find the diamond norm between $\mathcal{C}_X^{\otimes N}$ ($N$ copies of $\mathcal{C}_X$) and $\mathcal{C}_Y^{\otimes N}$. The only reason that such a lower bound would not be tight is if adaptivity between rounds adds to the discriminative power of a protocol (it is not yet known whether this is the case). The problem with using such a bound is that it is difficult to find the diamond norm for $N>1$ (as discussed in Subsection~\ref{subsection: diamond norm}).

An alternative is to consider specific protocols that could be implemented and to find the trace distances between outputs in these cases. Since we are looking for the maximum trace distance over all possible protocols, any specific protocol provides a lower bound on this maximum.

Ref.~\cite{pirandola_fundamental_2019} provided a lower bound on the trace norm between protocol outputs, based on consideration of a non-adaptive protocol, in which $N$ copies of a Bell state are sent through the channel. The output of this protocol is $N$ copies of the Choi matrix of the channel. It was shown that
\begin{align}
&D^{\mathrm{opt},N}_{\mathcal{C}_X \mathcal{C}_Y}\geq 2\left(1-f^{\mathrm{Choi}}(\eta_X,\eta_Y)^{N}\right),\\
&f^{\mathrm{Choi}}(p,q)=\frac{1+\sqrt{(1-p)(1-q)}+\sqrt{p q}}{2}.
\end{align}
$f^{\mathrm{Choi}}$ is the fidelity between the Choi matrices of channels $\mathcal{C}_X$ and $\mathcal{C}_Y$. We refer to this as the Bell state lower bound.

In fact, we find that we can obtain a slightly tighter bound using an alternative, non-adaptive protocol, in which $N$ copies of the state $\left|1\right>$ are sent through the channel. Note that this is also the input state that achieves the maximum quantum Fisher information (QFI) per channel use~\cite{adesso_optimal_2009}, and so is the optimal input state for parameter estimation, at least in the asymptotic limit of a large number of channel uses. In this case, we obtain the tighter bound
\begin{align}
&D^{\mathrm{opt},N}_{\mathcal{C}_X \mathcal{C}_Y}\geq 2\left(1-f^{\left|1\right>}(\eta_X,\eta_Y)^N\right),\label{eq: lower bound}\\
&f^{\left|1\right>}(p,q)=\sqrt{(1-p)(1-q)}+\sqrt{p q}.
\end{align}
This bound is again based on the fidelity between the possible outputs of the protocol.

For sufficiently small $N$, we can do better still by calculating the exact trace norm for this protocol (rather than lower bounding it). Since the output state of the protocol takes the form
\begin{align}
\rho_{X(Y)}^{N,\mathrm{out}}=\left((1-\eta_{X(Y)})\left|0\right>\left<0\right|+\eta_{X(Y)}\left|1\right>\left<1\right|\right)^{\otimes N},
\end{align}
for channel $\mathcal{C}_{X(Y)}$, the trace norm between the two possible outputs, $D^{\left|1\right>,N}_{\mathcal{C}_X \mathcal{C}_Y}$, is
\begin{align}
D^{\left|1\right>,N}_{\mathcal{C}_X \mathcal{C}_Y}=\sum_{i=0}^N \binom{N}{i}\left|\eta_{X}^{N-i}(1-\eta_X)^{i}-\eta_{Y}^{N-i}(1-\eta_Y)^{i}\right|.\label{eq: trace norm lower bound}
\end{align}
We will refer to this bound as the improved lower bound. The problem with using this bound for large $N$ is that the binomial coefficients become large, and therefore computationally difficult to calculate.

When applying the trace norm bounds to channel discrimination, the improved lower bound can be approximated using the quantum Cram\'{e}r-Rao bound, as per Ref.~\cite{spedalieri_detecting_2020}. The QCRB lower bounds the error-variance for estimating a channel parameter, based on the QFI with respect to that parameter. In our case, we have
\begin{align}
\sigma^2_{\eta}\leq\frac{1}{NH_{\eta}},
\end{align}
where $\sigma^2_{\eta}$ is the variance of an estimation of $\eta$ around its true value and $H_{\eta}$ is the QFI with respect to $\eta$. As shown in Ref.~\cite{adesso_optimal_2009}, the optimal QFI is achieved using number states with the maximum number of photons per channel use. In our case, this is the state $\left|1\right>$ and the maximum QFI per channel use is
\begin{align}
H_{\eta}^{\mathrm{max}}=\frac{1}{\eta(1-\eta)}.
\end{align}
The QCRB therefore takes the form
\begin{align}
\sigma^2_{\eta}\leq\frac{\eta(1-\eta)}{N}.\label{eq: QCRB}
\end{align}

We can return to a binary hypothesis testing scenario by picking a threshold value, $\tau$, of $\eta$, such that if our estimation of $\eta$ is greater than $\tau$, we decide that we have channel $\mathcal{C}_X$ (for $\eta_X>\eta_Y$), and if not, we decide that we have channel $\mathcal{C}_Y$. We assume that our estimation of $\eta$, $\eta'$, follows a Gaussian probability distribution, centred on the true value of $\eta$, with a variance equal to the lower bound from Eq.~(\ref{eq: QCRB}). For $\mathcal{C}_{X(Y)}$, this distribution is given by
\begin{align}
p_{\eta_{X(Y)}}(\eta')=\frac{1}{\sigma_{\eta}\sqrt{2\pi}}e^{-\frac{\left(\eta'-\eta_{X(Y)}\right)^2}{2\sigma^2_{\eta_{X(Y)}}}}.
\end{align}

Ref.~\cite{spedalieri_detecting_2020} then calculated the probabilities of deciding we have channel $\mathcal{C}_Y$ when we have channel $\mathcal{C}_X$ ($p_{X}^{\mathrm{err}}$) and of deciding we have channel $\mathcal{C}_X$ when we have channel $\mathcal{C}_Y$ ($p_{Y}^{\mathrm{err}}$). These error probabilities are
\begin{align}
&p_{X}^{\mathrm{err}}=\mathcal{N}_{X}^{-1}\int_{0}^{\tau} p_{\eta_{X}}(\eta')d\eta',\label{eq: QCRB err X}\\
&p_{Y}^{\mathrm{err}}=\mathcal{N}_{Y}^{-1}\int_{\tau}^{1} p_{\eta_{Y}}(\eta')d\eta',\label{eq: QCRB err Y}\\
&\mathcal{N}_{X(Y)}=\int_{0}^{1} p_{\eta_{X(Y)}}(\eta')d\eta',
\end{align}
where the normalisation factors, $\mathcal{N}_{X(Y)}$, are due to restricting the probability distributions to the range $[0,1]$, and again assuming $\eta_X>\eta_Y$. In the case in which both channels have prior probabilities, we can then choose the value of $\tau$ that minimises the mean of these two errors, in order to find the total error probability (in the asymmetric case, we can minimise a weighted mean of the errors). This then gives us an estimate of the error probability obtained using the improved lower bound on the trace norm.

It should be noted that this estimate is only tight for a large number of channel uses. For a small number of channel uses, the QCRB is often not tight~\cite{braunstein_maximum-likelihood_1992}. Since Eq.~(\ref{eq: QCRB}) lower bounds the variance of the parameter estimates (rather than upper bounding them), we do not attain an upper bound on the error probability, but rather an estimate of the error probability attained using the upper bound in Eq.~(\ref{eq: trace norm lower bound}) (since we are using the same input states in both cases, and the trace norm between the output states gives the lowest possible error in discriminating between them). In fact, for low $N$, the estimate of the bound based on the QCRB, which we will call the QCRB bound, underestimates the minimum error probability over a range of values. We will therefore only apply it for large $N$ ($>100$). The advantage of using it in this range is that it is more easily calculated than the upper bound in Eq.~(\ref{eq: trace norm lower bound}), whilst being significantly tighter than the bound on the error probability attained using Eq.~(\ref{eq: lower bound}).

\subsection{Upper bounds from PBT simulations}\label{subsection: PBT upper bounds}
We now calculate the upper bounds based on PBT simulations of the AD channel. We consider three types of resource state, as mentioned in Subsection~\ref{subsection: processors}.

The first type is the Choi matrix of the simulated channels, resulting in the upper bounds in Ref.~\cite{pirandola_fundamental_2019}, which we call the standard Choi bounds. These bounds are
\begin{align}
&D^{\mathrm{opt},N}_{\mathcal{C}_X \mathcal{C}_Y}\leq N\epsilon^{\mathrm{std}}_{m,XY}+2\sqrt{1-f^{\mathrm{Choi}}(\eta_X,\eta_Y)^{2mN}},\\
&\epsilon^{\mathrm{std}}_m(\eta)=\xi_m\left(\frac{\eta}{2}+\sqrt{\eta}\right),\\
&\epsilon^{\mathrm{std}}_{m,XY}=\epsilon^{\mathrm{std}}_m(\eta_X)+\epsilon^{\mathrm{std}}_m(\eta_Y),
\end{align}
where $m$ can take any positive, integer value. Note that we have a family of bounds, since we have a bound for any value of the number of ports, $m$. We must then optimise over $m$ to achieve the tightest possible bound in this family.

The second type of resource state is similar to the first, but $\eta_X$ and $\eta_Y$ have been replaced by $\eta'_X$ and $\eta'_Y$, according to Eq.~(\ref{eq: improved Choi eta}). The reason we choose this value of $\eta'_{X(Y)}$ is that this is one of the points at which the diamond norm between the channel and its simulation coincides with the trace norm. This means that we have an analytical expression for the resulting family of trace norm bounds, which we call the improved Choi bounds. Further, for all values of $m\geq 6$, the simulation errors are lower than for the standard Choi resource. These bounds are
\begin{align}
&D^{\mathrm{opt},N}_{\mathcal{C}_X \mathcal{C}_Y}\leq N\epsilon^{\mathrm{imp}}_{m,XY}+2\sqrt{1-f^{\mathrm{Choi}}(\eta'_X,\eta'_Y)^{2mN}},\\
\begin{split}
&\epsilon^{\mathrm{imp}}_m(\eta)=\frac{1}{2}\left(\frac{(\eta)\xi_m}{1-\xi_m}+\right.\\
&\; \; \; \; \; \; \; \; \left.\sqrt{4(\eta)\left(1-\sqrt{1-\xi_m}\right)^2+\frac{\eta^2\xi_m^2}{(1-\xi_m)^2}}\right),
\end{split}\\
&\epsilon^{\mathrm{imp}}_{m,XY}=\epsilon^{\mathrm{imp}}_m(\eta_X)+\epsilon^{\mathrm{imp}}_m(\eta_Y).
\end{align}

In fact, since the chosen values of $\eta'_X$ and $\eta'_Y$ are not necessarily the values that give the tightest possible bounds, we could numerically minimise over all pairs of ``Choi-like" resources simulating $\mathcal{C}_X$ and $\mathcal{C}_Y$. In other words, we could simulate $\mathcal{C}_X$ with $R^{\mathrm{Choi}}(\eta''_X)$ and $\mathcal{C}_Y$ with $R^{\mathrm{Choi}}(\eta''_Y)$, where
\begin{align}
&R^{\mathrm{Choi}}(\eta)=C(\eta)^{\otimes m},\\
&C(\eta)=\begin{pmatrix}
\frac{1-\eta}{2} &0 &0 &0\\
0 &\frac{1}{2} &-\frac{\sqrt{\eta}}{2} &0\\
0 &-\frac{\sqrt{\eta}}{2} &\frac{\eta}{2} &0\\
0 &0 &0 &0
\end{pmatrix}.
\end{align}
Note that, in the context of ``Choi-like" resources, the Bell pair used to generate the Choi state is the state $\frac{1}{\sqrt{2}}(\left|01\right>-\left|10\right>)$. We could then numerically minimise over $\eta''_X$ and $\eta''_Y$ to find the optimal resource states. However, in this case, we would not have an analytical expression for the simulation error, and would need to calculate it numerically, by finding the diamond norm for both simulations; this involves maximising the trace norm between the channels and their simulations over all possible input states. We would then also need to minimise the bounds over $m$. This would involve a lot more numerical minimisation/maximisation than simply numerically optimising the standard and improved Choi bounds over $m$. This is why we do not find optimal bounds for this more general resource.

Finally, we consider resource states, $R^{\mathrm{alt}}(a)$, of the form given in Eq.~(\ref{eq: alt resource}). The Choi matrix of the channel simulated by carrying out PBT with this resource state is given in Ref.~\cite{pereira_characterising_2021}. Writing the Choi matrix in the form
\begin{align}
\rho^{\mathrm{Choi}}_{\mathrm{PBT}(R^{\mathrm{alt}}(a))}=\begin{pmatrix}
x &0 &0 &z\\
0 &\frac{1}{2}-x &0 &0\\
0 &0 &y &0\\
z &0 &0 &\frac{1}{2}-y
\end{pmatrix},
\end{align}
where the expressions for $x$, $y$ and $z$ are functions of $a$ (and $m$), which are given in Ref.~\cite{pereira_characterising_2021}, we choose $a_{X(Y)}$ such that
\begin{align}
x\left(a_{X(Y)}\right)-y\left(a_{X(Y)}\right)=\frac{\eta_{X(Y)}}{2}.
\end{align}
We then simulate $\mathcal{C}_{X(Y)}$ with $a_{X(Y)}$, and call the resulting bounds the alternative resource bounds. We choose this value of $a_{X(Y)}$ because this is one of the points at which the diamond norm between the channel and its simulation coincides with the trace norm. Similarly to the case of the ``Choi-like" resources, we could minimise our bound over all possible values of $a_{X(Y)}$, rather than choosing this value, but this would again require a lot more numerical minimisation/maximisation. The simulation error is given by
\begin{align}
\begin{split}
\epsilon^{\mathrm{alt}}(a&_{X(Y)})=1-\eta_{X(Y)}-2y+\\
&\left.\sqrt{(1-\eta_{X(Y)}-2y)^2+(\sqrt{\eta_{X(Y)}}-2z)^2}\right|_{a=a_{X(Y)}},
\end{split}
\end{align}
and the fidelity between the resource states is given by
\begin{align}
f^{\mathrm{alt}}(a_X,a_Y)=\sqrt{a_X a_Y}+\sqrt{(1-a_X)(1-a_Y)}.
\end{align}
The alternative resource bounds are therefore given by
\begin{align}
&D^{\mathrm{opt},N}_{\mathcal{C}_X \mathcal{C}_Y}\leq N\epsilon^{\mathrm{alt}}_{XY}+2\sqrt{1-f^{\mathrm{alt}}(a_X,a_Y)^{2mN}},\\
&\epsilon^{\mathrm{alt}}_{XY}=\epsilon^{\mathrm{alt}}(a_X)+\epsilon^{\mathrm{alt}}(a_Y).\label{eq: alt resource bound}
\end{align}
Although this may not be immediately apparent from the expressions, both $\epsilon^{\mathrm{alt}}_{XY}$ and $f^{\mathrm{alt}}$ depend on $m$, since the expressions for $x$, $y$, $z$ and $a_{X(Y)}$ all depend on $m$. Therefore, we again want to pick the optimal value of $m$, in order to achieve the tightest possible bound. Note that the resource states are pure, meaning that our expression for the trace norm between different resource states is exact.

In order to optimise over $m$, we use analytical functions that closely approximate the standard and improved Choi bounds, but that do not feature $\xi_m$. This is done because the expression for $\xi_m$ is too complicated to easily find an analytical minimum of the full bounds. Specifically, we replace $\xi_m$ in the simulation error expressions with $m^{-1}$; this gives us expressions that we can easily locate the minima of, for fixed $\eta_X$, $\eta_Y$ and $N$. We then use the closest integer values of $m$ to our minima when calculating the actual values of the bounds (substituting them into the original expressions). When referring to the standard or improved Choi bound in Section~\ref{section: numerical}, it is implicit that this process has been carried out, and that the bounds are calculated for the optimal value of $m$. For the alternative resource bounds, we find numerically that the bound gets tighter as $m$ increases, rather than having a maximum, so we pick a fixed, high value of $m$.

\subsection{Extending to the qudit case}\label{subsection: qudit}
We will briefly consider the case in which we must discriminate between two pure loss, qudit channels, rather than two AD channels (which are pure loss, qubit channels). The Stinespring dilation of such a channel is a beamsplitter acting on an environmental vacuum mode. The action of the beamsplitter can be described as
\begin{align}
\left|n\right>_{S}\left|0\right>_{E}\to\sum_{i=0}^n \sqrt{\eta^{n-i}(1-\eta)^i\binom{n}{i}}\left|n-i\right>_{S}\left|i\right>_{E},
\end{align}
where $S$ labels the signal mode (the input mode to the channel), $E$ labels the environmental mode and $\eta$ is the transmissivity of the beamsplitter (and the channel). The binomial coefficient on the right-hand side of the expression comes from the choice of which photons are transferred to the environmental modes. This means that a $d$-dimensional, pure loss channel, with transmissivity $\eta$, can be described by the $d$ Kraus operators
\begin{align}
K_j=\sum_{i=j}^{d-1} \sqrt{\eta^{i-j}(1-\eta)^{j}\binom{i}{j}}\left|i\right>\left<i-j\right|,
\end{align}
where the label $j$ ranges from $0$ to $d-1$.

Calling our pure loss, $d$-dimensional channels $\mathcal{C}_{X}^d$ and  $\mathcal{C}_{Y}^d$, with transmissivities of $\eta_X$ and $\eta_Y$ respectively, the $J$-matrix of the two channels (the difference between the Choi matrices, multiplied by the input dimension of the channel, as per Subsection~\ref{subsection: diamond norm}) can be written as
\begin{align}
&J(\mathcal{C}_X^d,\mathcal{C}_Y^d)=\sum_{i=0}^{d-1} \left(\left|v_{\eta_X}^i\right>\left<v_{\eta_X}^i\right| - \left|v_{\eta_Y}^i\right>\left<v_{\eta_Y}^i\right|\right)_{SI},\label{eq: difference matrix qudit}\\
&\left|v_{\eta}^i\right>_{SI}=\sum_{j=0}^{d-i-1}\sqrt{\eta^{j}(1-\eta)^{i}\binom{i+1}{j}}\left|i+j\right>_S\left|j\right>_I,
\end{align}
where $S$ labels the signal mode and $I$ labels the idler mode.

In order to calculate simulation bounds in the qudit case, we would require expressions for the output of qudit PBT channels, and would require new resource states capable of simulating pure loss, qudit channels. However, the trivial bound, based on the diamond norm, can still be used in the qudit case (substituting the diamond norm between channels $\mathcal{C}_{X}$ and  $\mathcal{C}_{Y}$, in Eq.~(\ref{eq: trivial bound}), with the diamond norm between $\mathcal{C}_{X}^d$ and  $\mathcal{C}_{Y}^d$).

We do not have an analytical expression for the diamond norm between channels $\mathcal{C}_{X}^d$ and  $\mathcal{C}_{Y}^d$, as we do for the qubit case. Instead, we can find it numerically, using semidefinite programming and the formula for the difference between Choi matrices, given in Eq.~(\ref{eq: difference matrix qudit}). The issue here is the same as with finding the diamond norm for multiple channel uses. As the input dimension becomes large (i.e. for large $d$), numerically finding the diamond norm becomes computationally expensive. An alternative is to bound the diamond norm using a result from Ref.~\cite{nechita_almost_2018}.

Ref.~\cite{nechita_almost_2018} showed that the diamond distance between any two channels, $\mathcal{A}$ and $\mathcal{B}$, can be bounded by
\begin{align}
\left\|\mathcal{A}-\mathcal{B}\right\|_{\diamond}\leq \left\|\mathrm{Tr}_{S}\left|J(\mathcal{A},\mathcal{B})\right|\right\|_{\infty},
\end{align}
where $J(\mathcal{A},\mathcal{B})$ is the difference between the Choi matrices of $\mathcal{A}$ and $\mathcal{B}$, multiplied by the input dimension of the channels. Here we have first taken the absolute value of the matrix $J(\mathcal{A},\mathcal{B})$, then taken the partial trace over the signal mode. We have then taken the largest eigenvalue (the operator norm) of the resulting matrix as our bound. Note that this coincides with the trace norm, and is therefore exactly the diamond distance, if the matrix is scalar 
after the partial trace is taken. Applying this bound to the expression in Eq.~(\ref{eq: difference matrix qudit}) gives a computationally cheaper (but less tight) bound on the diamond norm (and hence on the optimal trace distance between protocol outputs) than finding the diamond norm numerically, via semidefinite programming.

Numerical investigation shows that the diamond norm between two qudit channels, $\mathcal{C}_{X}^d$ and  $\mathcal{C}_{Y}^d$, appears to coincide with the diamond norm between $d-1$ uses of two qubit channels, $\mathcal{C}_{X}^{\otimes d-1}$ and  $\mathcal{C}_{Y}^{\otimes d-1}$ (for the same transmissivities, $\eta_X$ and $\eta_Y$). This suggests some connection between the two cases, however it is not clear what the connection is.

It is also worth noting that the (approximation of the) upper bound on the error probability of discriminating between two equiprobable channels attained by using the QCRB still holds in the qudit case (as long as the number of channels is large enough for the approximation to be valid), with the only change being to the lower bound on the channel parameter variance, in Eq.~(\ref{eq: QCRB}). In this equation, $N(d-1)$ is substituted for $N$. This is because the maximum QFI per channel use is
\begin{align}
H^{\mathrm{max}}_{\eta,d}=\frac{d-1}{\eta(1-\eta)},
\end{align}
and the input state that attains this value is $\left|d-1\right>$~\cite{adesso_optimal_2009,spedalieri_detecting_2020}. The QFI is additive, so the maximum value of the total QFI is the same for both lossy channels and AD channels, as long as the total number of photons sent through the channel is the same.

\section{Numerical investigations}\label{section: numerical}
Carrying out numerical PBT simulations for our three classes of resource states over a variety of $\eta_X$, $\eta_Y$ and $N$ values, we find that the improved Choi bound beats the standard Choi bound over the entire range of investigated parameter values. We also find that the alternative resource bound, for a sufficiently large number of ports, $m$, beats the both Choi bounds across almost the entire range, and that the trivial bound also beats the Choi bounds over a wide range of values. In fact, either the trivial bound or the alternative resource bound beat both of the Choi bounds over the entire range of values that was investigated. Since this was a numerical study, it is not possible to definitively say that the tightest out of the trivial bound and the alternative resource bound is always tighter than either of the Choi bounds, however this is the case for a wide range of parameter values.

\begin{figure}[tb]
\centering
\includegraphics[width=1\linewidth]{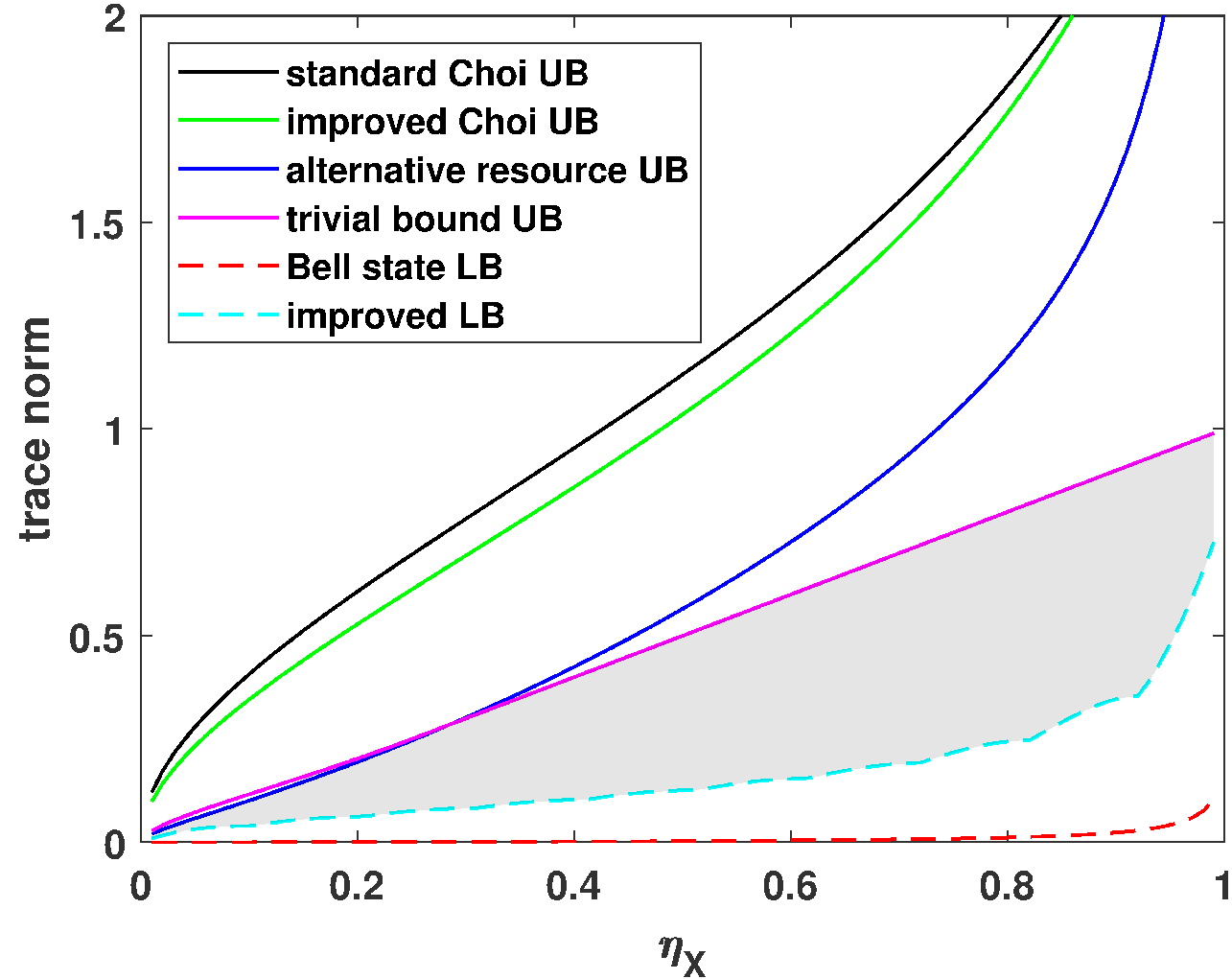}\caption{Upper and lower bounds on the maximum value of the trace norm between the two possible outputs of an adaptive discrimination protocol with no more than 10 channel uses. The channels being discriminated between are AD channels with transmissivities $\eta_X$ and $\eta_Y$, where $\eta_Y=\eta_X\eta_{XY}$. In this case, $\eta_{XY}=0.95$. The two upper bounds based on PBT simulations using ``Choi-like" resources are significantly less tight than the trivial (upper) bound and the upper bound based on PBT simulations using the alternative resource. Each of these latter two bounds is optimal over some range of $\eta_X$ values. The improved lower bound is tighter than the Bell state lower bound. The grey shaded area is the region between the tightest upper and lower bounds.}
\label{fig: bound comp, eta = 0.95}
\end{figure}

\begin{figure}[tb]
\centering
\includegraphics[width=1\linewidth]{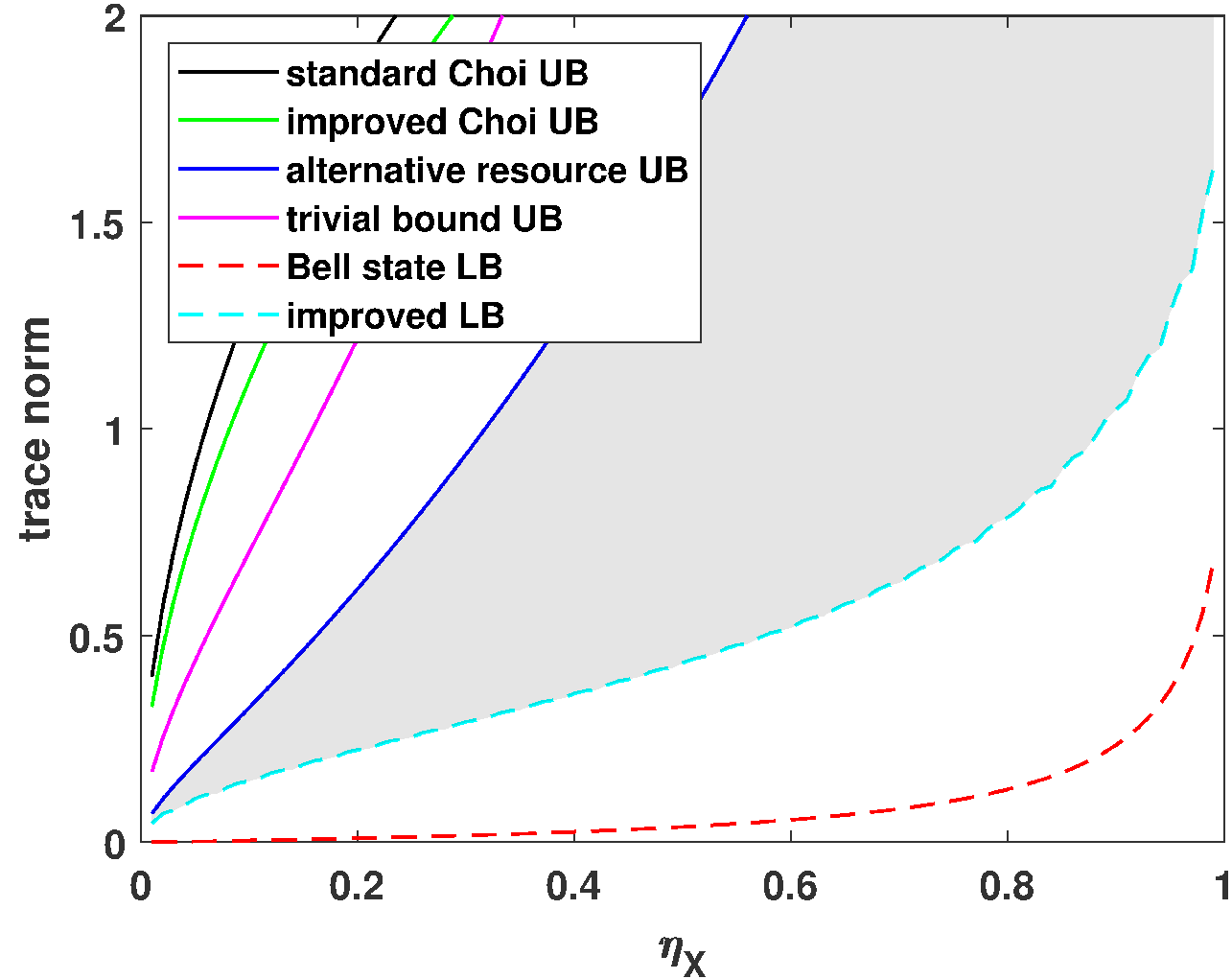}\caption{Upper and lower bounds on the maximum value of the trace norm between the two possible outputs of an adaptive discrimination protocol with no more than 30 channel uses. The channels being discriminated between are AD channels with transmissivities $\eta_X$ and $\eta_Y$, where $\eta_Y=\eta_X\eta_{XY}$. In this case, $\eta_{XY}=0.9$. For these parameter values, the upper bound based on simulation using the alternative resource is always better than the other three upper bounds. It is to be expected that the trivial bound performs less well for high values of $N$, because it scales linearly with $N$, whilst the bounds based on PBT do not. The improved lower bound has a distinct advantage over the Bell state lower bound. The grey shaded area is the region between the tightest upper and lower bounds.}
\label{fig: bound comp, eta = 0.9}
\end{figure}

\begin{figure}[tb]
\centering
\includegraphics[width=1\linewidth]{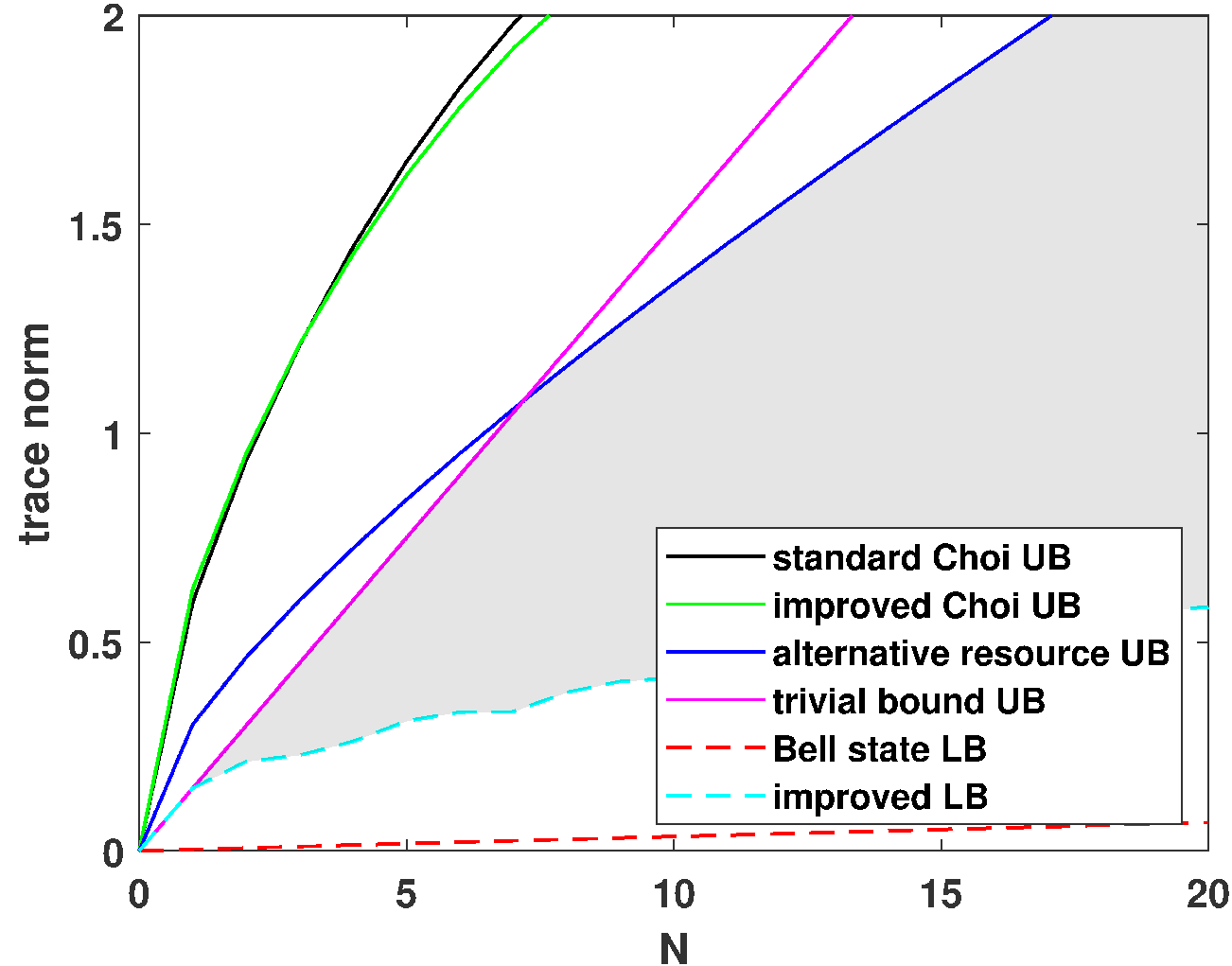}\caption{Upper and lower bounds on the maximum value of the trace norm between the two possible outputs of an adaptive discrimination protocol for varying numbers of channel uses. The channels being discriminated between are AD channels with transmissivities $\eta_X=0.75$ and $\eta_Y=0.675$. For low $N$, the trivial bound is the tightest upper bound, whilst for $N>7$, the alternative resource bound is the tightest upper bound. The grey shaded area is the region between the tightest upper and lower bounds.}
\label{fig: bound comp N}
\end{figure}

In Figs.~\ref{fig: bound comp, eta = 0.95} and \ref{fig: bound comp, eta = 0.9}, we demonstrate the performance of the various bounds. Choosing $\eta_X>\eta_Y$, we decompose $\mathcal{C}_Y$ as the pointwise application of $\mathcal{C}_X$ and some other AD channel, $\mathcal{C}_{XY}$, with transmissivity
\begin{align}
\eta_{XY}=\frac{\eta_Y}{\eta_X}.
\end{align}
Two specific values of $\eta_{XY}$ (one per plot) were chosen: $0.95$ and $0.9$. Two values of $N$, the total number of channel uses, were also chosen: $10$ and $30$. With these kept fixed, $\eta_X$ was then varied from $0.01$ to $0.99$ and the bounds were studied over this range. For the alternative resource bound, we have set $m=150$.

As shown in the plots, the improved Choi bound performs better than the standard Choi bound, however both are beaten by either the trivial bound or the alternative resource bound (which of these is highest depends on the parameter values). The trivial bound performs better than the alternative resource bound when $\eta_X$ and $\eta_{XY}$ are large and when $N$ is small.

Fig.~\ref{fig: bound comp N} illustrates the $N$-dependence of the various trace norm bounds. In this plot, $\eta_X=0.75$ and $\eta_{XY}=0.9$. The value of $N$ is then varied from $0$ to $20$. The trivial bound (which is tight for $N=1$) is the tightest upper bound for low $N$, but is beaten by the alternative resource bound for $N>7$, because the former scales linearly whilst the latter scales sub-linearly.

The new lower bound on the optimal trace distance (based on sending $N$ copies of the state $\left|1\right>$ through the channel) is tighter than the lower bound from Ref.~\cite{pirandola_fundamental_2019} (based on sending $N$ copies of a Bell state through the channel) across the entire range. It is clear, however, that there is still room to tighten either the upper or the lower bound, since there is still a gap between the tightest upper bound and the tightest lower bound, especially in Fig.~\ref{fig: bound comp, eta = 0.9}.

\begin{figure}[tb]
\centering
\includegraphics[width=1\linewidth]{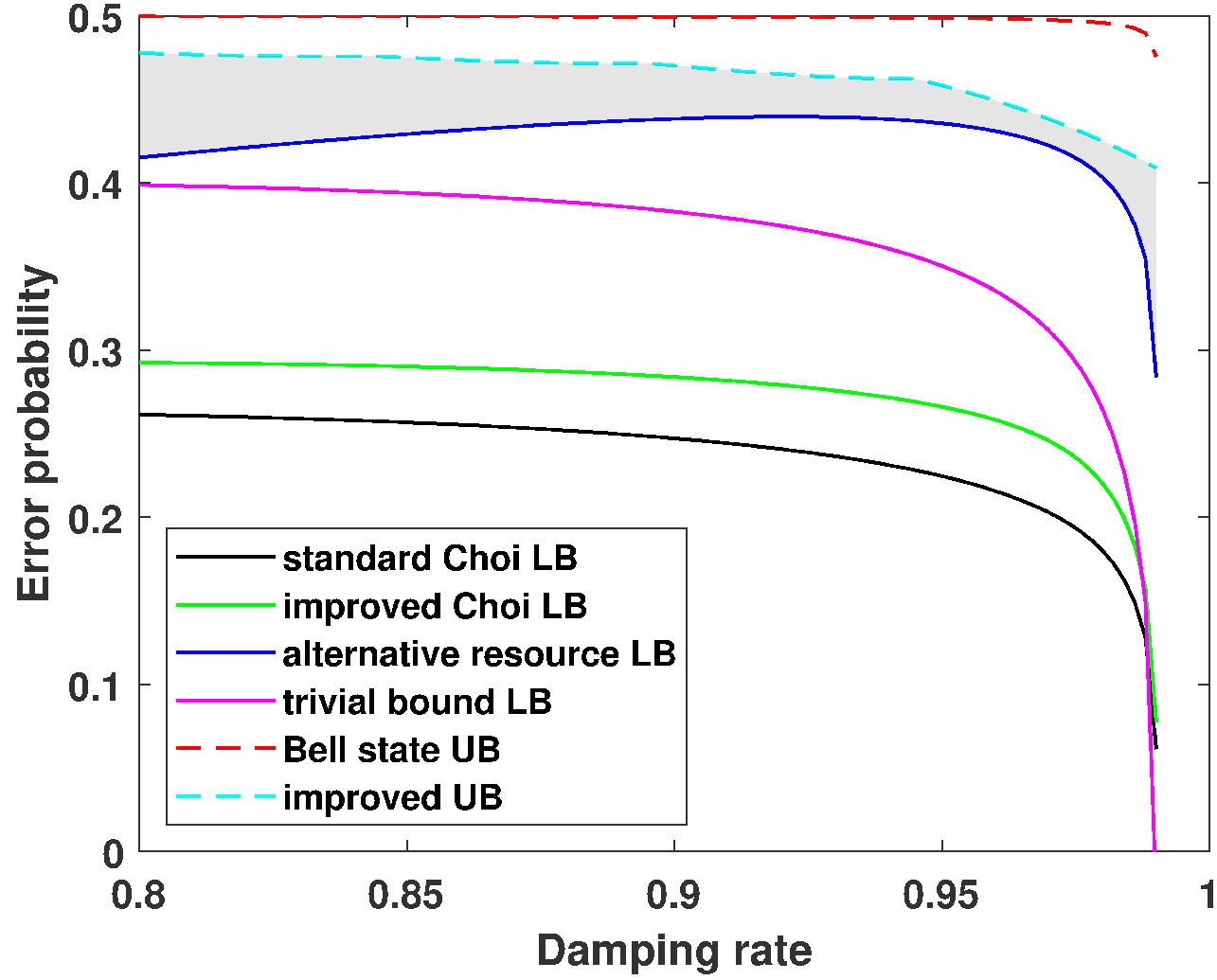}\caption{Comparison with the bounds on the error probability of discriminating between two AD channels, one with damping rate $p$ and one with damping rate $p+0.01$, with equal prior probabilities, with no more than $20$ channel uses, found in Ref.~\cite{pirandola_fundamental_2019}. The line labelled ``standard Choi" is the lower bound found in Ref.~\cite{pirandola_fundamental_2019}, and the line labelled ``Bell state UB" is the upper bound from the same paper. The other lines are the new bounds presented here, and the grey shaded area is the region between the tightest upper and lower bounds.}
\label{fig: paper comparison}
\end{figure}

In Fig.~\ref{fig: paper comparison}, we compare our new bounds to the bounds presented in Fig.~4 of Ref.~\cite{pirandola_fundamental_2019}. We plot the error probability of discriminating between two AD channels, with equal prior probability, using the equation in Eq.~(\ref{eq: error probability}). The AD channels are characterised by the damping rate, $p$, rather than by the transmittance, $\eta$, although the two quantities are trivially connected via the equation $p=1-\eta$. One channel has a damping rate of $p$ and the other has a damping rate of $p+0.01$. The maximum number of channel uses is $20$. The new lower bounds on the error probability (which come from the new upper bounds on the trace norm) are tighter than the lower bounds in Ref.~\cite{pirandola_fundamental_2019} over the entire range; in this case, the alternative resource bound is the tightest lower bound. The new upper bound on the error probability (coming from the improved lower bound on the optimal trace norm) is slightly tighter than the upper bound in Ref.~\cite{pirandola_fundamental_2019} over the entire range, but most noticeably for a high damping rate, $p$.

We now consider two examples of how these bounds might be applied to quantum information tasks. The tasks we consider are quantum hacking and biological sensing (a quantum metrology task).

\subsection{Applying the bounds to quantum hacking}\label{subsection: quantum hacking}
Suppose a hacker, Eve, is attempting to eavesdrop on communications between a sender, Alice, and a receiver, Bob, who are implementing the BB84 protocol. Suppose also that Eve is able to send photons into Alice's device before each transmission (and to receive some return state). Eve could use this side-channel to gain more information on the states sent by Alice than is accounted for in the security proofs. For instance, Alice's basis choice could be enacted by a polariser~\cite{kim_implementation_2008}. By sending in photons with a known polarisation, Eve could glean information about Alice's basis choice based on the loss experienced by the photons (which could be basis dependent). Then, if Eve could determine Alice's basis with a high probability of success, she could carry out an intercept and resend attack on the photons sent through the main channel, without greatly disturbing them. In other words, she could measure the signal states in the basis that she believes them to have been sent in, based on her side-channel attack. Alice and Bob would only detect errors in half of the cases in which Eve incorrectly guesses Alice's basis. Since, in this scenario, Eve's error probability is low, the quantum bit error rate (QBER) detected by the trusted parties would be much lower than the $25\%$ normally expected for an intercept and resend attack.

We model the attack as Eve carrying out a general, adaptive discrimination protocol with up to $N$ channel uses. We set the transmissivity of one of the channels as $\eta_Y=0$, and then choose three different values of $\eta_X$: $10^{-5}$, $5\times10^{-6}$ and $10^{-6}$. We then calculate Eve's discrimination error probability, assuming equal prior probabilities of each channel occurring, for various numbers of channel uses. In this scenario, we assume that we have a perfect polariser, and so for one channel (i.e. for one polarisation), the photons sent through are completely absorbed by the polariser, whilst for the other channel, they are undisturbed by the polariser. We assume that the input states are so strongly attenuated that they can be modelled as a train of at most single photon states by the time they arrive at the polariser, and hence that Eve's protocol can be modelled as a discrimination protocol between AD channels. This is a reasonable assumption, since BB84 involves the sending of single-photon states, which are often produced using strongly attenuated laser pulses. It is thus reasonable to assume that a laser pulse sent by Eve into Alice's device, through the optical fibre, would be similarly attenuated, such that the pulse arriving at the polariser could be well-modelled as a qubit state. We also assume that further attenuation occurs as the states leave the device, giving rise to the low values of $\eta_X$. This is in line with the architecture in Ref.~\cite{lucamarini_practical_2015}, which limits the total mean photon number leaving Alice's device via the optical fibre, per signal sent through the main channel, to $10^{-6}$.

The assumption that Eve's states can be modelled as (up to) one-photon states probing AD channels can be justified by numerically finding the energy-constrained diamond norm~\cite{pirandola_fundamental_2017,Shirokov,winter_energy-constrained_2017} between a lossy channel and the pointwise application of a truncation channel (a channel mapping all number states of the form $\left|n>1\right>$ to $\left|0\right>$) and the same lossy channel, for low transmissivities. More specifically, we use the semidefinite programme for calculating the energy-constrained diamond norm given in Ref.~\cite{winter_energy-constrained_2017}; note that the definition of the energy-constrained diamond norm used by Ref.~\cite{winter_energy-constrained_2017} (and Ref.~\cite{Shirokov}) differs slightly from the definition given by Ref.~\cite{pirandola_fundamental_2017}. We find that, for $\eta_X=10^{-6}$, the truncation to one-photon states has a small effect on the error probability. See the Supplementary data for further details~\cite{DATA}.

Since one of the channels ($\mathcal{C}_Y$) will always output the state $\left|0\right>$, we can significantly simplify the improved lower bound. Eq.~(\ref{eq: trace norm lower bound}) reduces to
\begin{align}
D^{\left|1\right>,N}_{\mathcal{C}_X \mathcal{C}_Y}=2(1-\eta_X^N).
\end{align}

\begin{figure}[tb]
\centering
\includegraphics[width=1\linewidth]{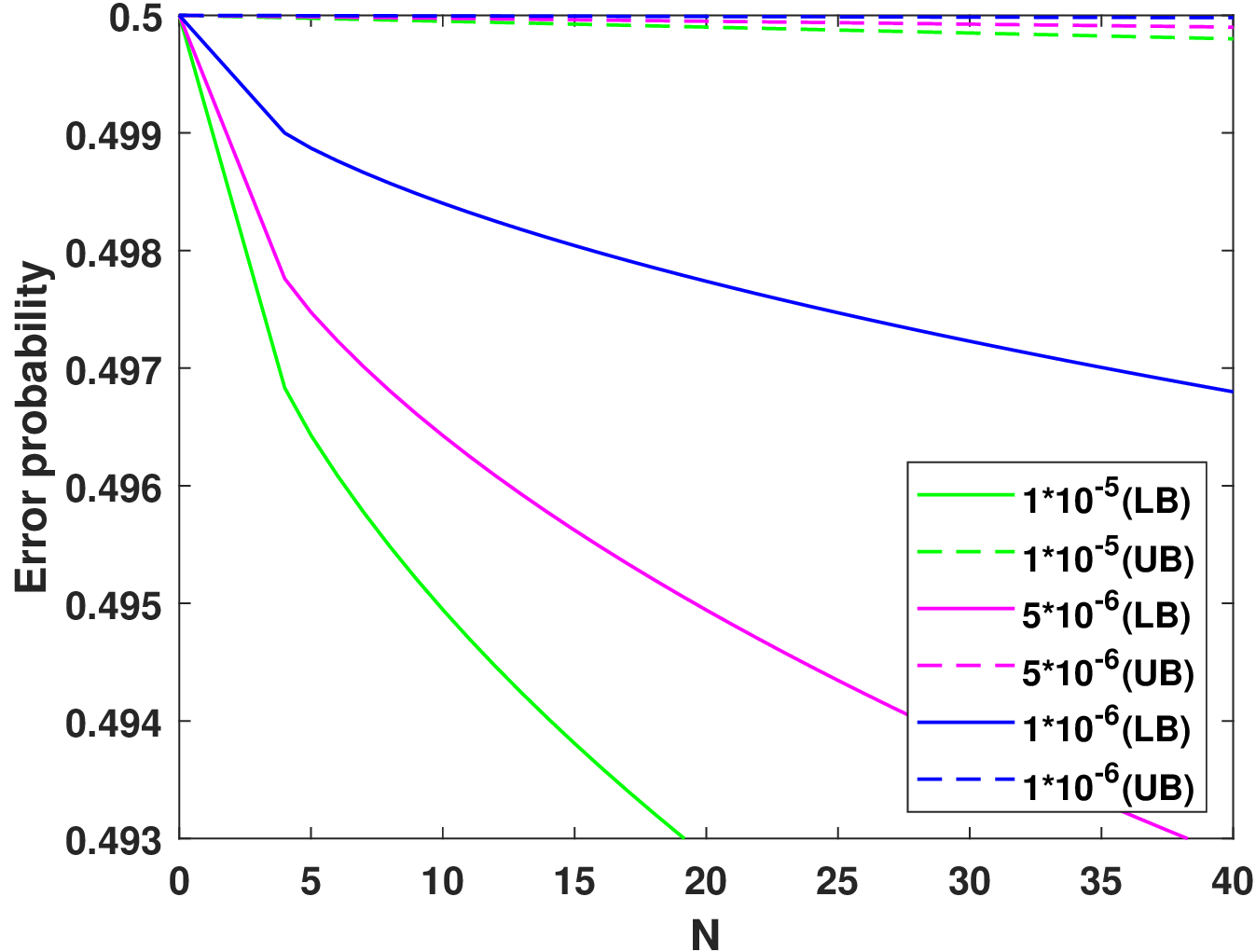}\caption{Upper and lower bounds on the discrimination error probability for an eavesdropper carrying out an adaptive protocol to discriminate between two BB84 preparation bases, with at most $N$ channel uses. We assume that Eve must send qubit states through an AD channel, in order to determine whether the channel has a transmissivity of $\eta_X$ or of $\eta_Y$. $\eta_Y=0$, whilst $\eta_X$ takes values of $10^{-5}$, $5\times10^{-6}$ and $10^{-6}$; each case is represented by a different colour. The continuous lines give lower bounds on the error probability, whilst the dashed lines give upper bounds. The upper bounds are based on the improved lower bound, from Eq.~(\ref{eq: trace norm lower bound}). The lower bounds are based on the trivial bound, from Eq.~(\ref{eq: trivial bound}), and the alternative resource bound, from Eq.~(\ref{eq: alt resource bound}); whichever bound has a higher value for a given $N$ is used for that value of $N$. For the alternative resource bound, $m=150$. We find that, for all three values of $\eta_X$, the trivial bound gives a tighter bound for $N\leq4$ and the alternative resource bound gives a tighter bound for $N>4$.}
\label{fig: hacking bounds}
\end{figure}

The upper and lower bounds found are shown in Fig.~\ref{fig: hacking bounds}. The upper bounds come from the improved lower bound and the lower bounds are based on whichever is tighter of the trivial bound and the alternative resource bound. For the chosen values of $\eta_X$, the trivial bound is tighter for $N\leq4$. This is in line with our expectation that the trivial bound performs less well (compared to bounds based on PBT simulation) for large values of $N$, due to its linear scaling. The gap between the upper and lower bounds is small in proportion to their values, but still shows significant room for improvement, especially for large $N$. It is not clear whether it is the upper bounds, the lower bounds, or both which need tightening.

\subsection{Applying the bounds to biological sensing}\label{subsection: biological sensing}
Quantum channel discrimination protocols have applications in biology. The concentration of bacteria in a growth medium affects the transmissivity of light through the medium. The tasks of distinguishing between the presence and absence of bacteria in a sample and of distinguishing between two possible concentrations of bacteria can therefore be considered to be quantum channel discrimination tasks, where the two possible channels are lossy channels with different transmissivities. Further, in biological applications, low photon numbers are often desirable, since intense radiation can harm the samples that are being probed. As a result, in some scenarios, modelling the task as an AD channel discrimination task may be appropriate.

In Ref.~\cite{spedalieri_detecting_2020}, the authors showed that quantum light sources and detectors can reduce the error probability for both detecting the presence or absence of \textit{E. coli} in a sample and determining whether a sample contains \textit{E. coli} or \textit{Salmonella}. They started by determining the transmissivities of growth media containing \textit{E. coli} and \textit{Salmonella} bacteria, as a function of time. The time-dependence comes from the changing concentrations of the bacteria in the media as they grow. The two possible types of bacteria and the case with no bacteria present therefore correspond to three different possible lossy channels. Determining whether a specific bacteria is present or absent and determining which of the two types of bacteria is present then become channel discrimination tasks.

Ref.~\cite{spedalieri_detecting_2020} considers the task of parameter estimation, where the parameter to be estimated is the transmissivity of the channel. It considers both coherent state sources and the optimal input states for parameter estimation from Ref.~\cite{adesso_optimal_2009} (which are number states that send the maximum number of photons through the channel per channel use). It then bounds the error probability for detecting the presence of \textit{E. coli} and discriminating between \textit{E. coli} and \textit{Salmonella}, by using the expressions in Eqs.~(\ref{eq: QCRB err X}) and~(\ref{eq: QCRB err Y}). In the symmetric testing case (equal prior probabilities), the mean of $p^{\mathrm{err}}_X$ and $p^{\mathrm{err}}_Y$ is minimised over $\tau$. Note, however, that the resulting expression (the QCRB bound) only provides an upper bound on the optimal error probability for sufficiently large $N$ (i.e. in the regime in which the QCRB is tight).

\begin{figure}[tb]
\centering
\includegraphics[width=1\linewidth]{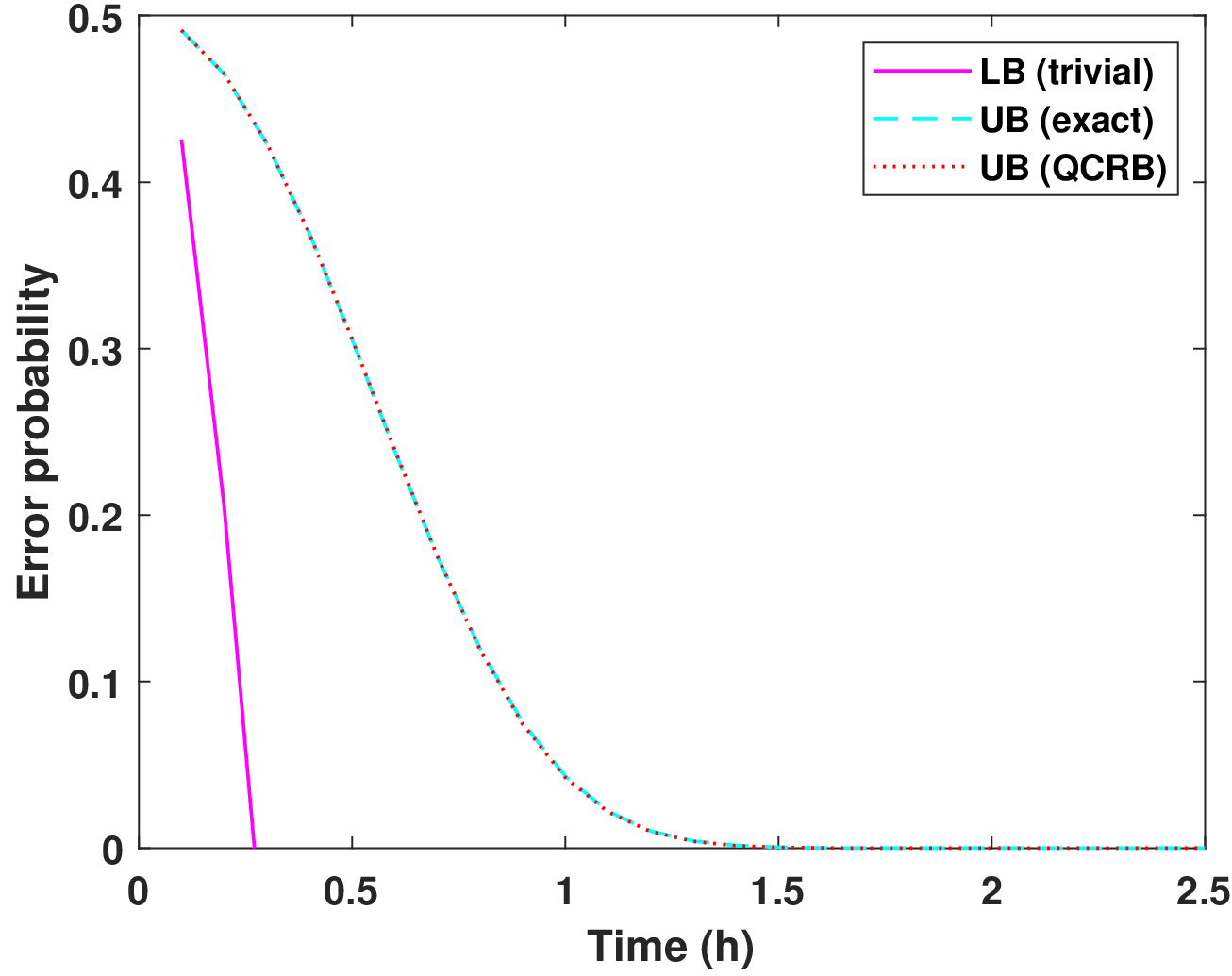}\caption{Upper and lower bounds on the error probability of detecting the presence of \textit{E. Coli} bacteria in a sample, with a maximum of $150$ channel uses (each using no more than one photon) as a function of time. The transmissivity of the blank sample is constant, whilst the transmissivity of the sample containing \textit{E. Coli} is modelled as following a cubic equation (with respect to the time since the sample was prepared). The lower bound (denoted ``LB (trivial)") is derived from the trivial bound on the trace norm. The exact form of the upper bound (``UB (exact)") is derived from the improved lower bound on the trace norm and the approximation to the upper bound (``UB (QCRB)") is based on the QCRB bound. Since the two bounds overlap almost perfectly, the approximation is valid in this regime.}
\label{fig: bacteria detection}
\end{figure}

\begin{figure}[tb]
\centering
\includegraphics[width=1\linewidth]{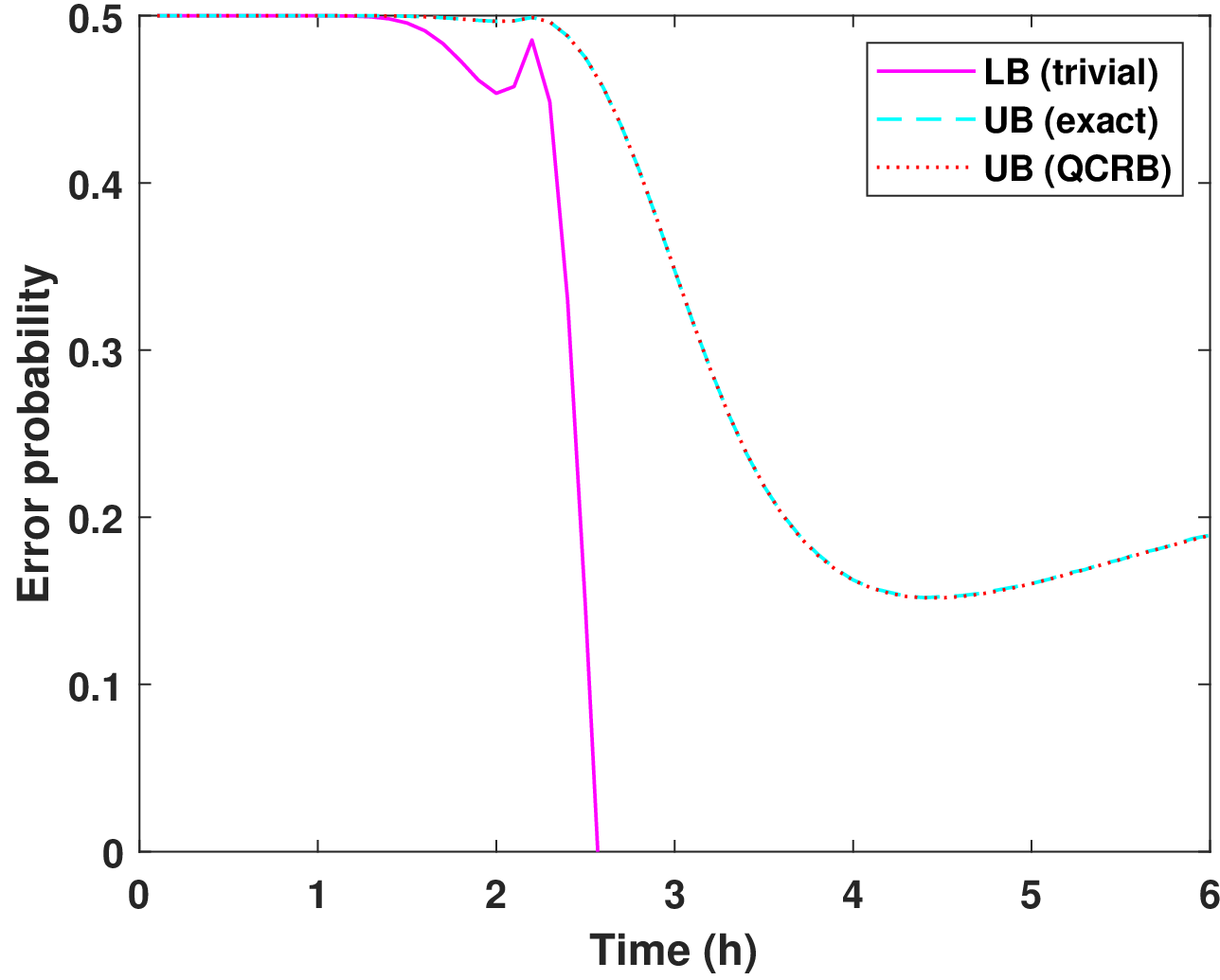}\caption{Upper and lower bounds on the error probability of discriminating between \textit{E. Coli} and \textit{Salmonella} bacteria in a sample, with a maximum of $150$ channel uses (each using no more than one photon) as a function of time. The absorbances of the samples are modelled as following Gompertz functions. The lower bound (denoted ``LB (trivial)") is derived from the trivial bound on the trace norm. The exact form of the upper bound (``UB (exact)") is derived from the improved lower bound on the trace norm and the approximation to the upper bound (``UB (QCRB)") is based on the QCRB bound. Since the two bounds overlap almost perfectly, the approximation is valid in this regime. The absorbances are initially very similar, but become more distinguishable as the time since the sample was prepared increases. We note that this plot differs from Fig.~10 in Ref.~\cite{spedalieri_detecting_2020}; this is because Ref.~\cite{spedalieri_detecting_2020} considers probing with a mean total of $10^3$ photons, whilst we only allow a maximum of $150$ photons in total. Ref.~\cite{spedalieri_detecting_2020} also models the transmissivities of the two samples using cubic equations, rather than Gompertz functions.}
\label{fig: bacteria discrimination}
\end{figure}

In Figs.~\ref{fig: bacteria detection} and \ref{fig: bacteria discrimination}, we plot upper and lower bounds on the optimal error probability for an adaptive protocol with up to $150$ channel uses, each sending at most one photon through the channel. This is reasonable, because it is desirable to send only a small amount of energy through the channels and because the error probabilities from Eqs.~(\ref{eq: QCRB err X}) and~(\ref{eq: QCRB err Y}) can be achieved in this way.

Fig.~\ref{fig: bacteria detection} bounds the error probability over time for detecting the presence of \textit{E. coli} in a sample. In this scenario, $\mathcal{C}_X$ is the channel corresponding to a blank sample (no bacteria present), and so $\eta_X$ has a constant value of $\eta_{\mathrm{bk}}=0.92$. $\mathcal{C}_Y$ is the channel corresponding to a sample with \textit{E. coli} present, and has a transmissivity of
\begin{align}
\eta_Y=\eta_{E.Coli}(t)=\eta_{bk}-c_{1,E.Coli}t^2+c_{2,E.Coli}t^3,\label{eq: E. coli cubic}
\end{align}
where $c_{1,E.Coli}$ and $c_{2,E.Coli}$ are constants with values of $0.1~\mathrm{h}^{-2}$ and $0.0088~\mathrm{h}^{-3}$ respectively and where $t$ is the time, in hours, since the sample was prepared. The values of $c_{1,E.Coli}$ and $c_{2,E.Coli}$ were experimentally determined in Ref.~\cite{spedalieri_detecting_2020}, and the cubic expression for $\eta_{E.Coli}$, from Eq.~(\ref{eq: E. coli cubic}) is valid for small $t$ ($\leq3$).

The lower bound is the tightest out of the lower bounds derived from our upper bounds on the trace norm. In fact, this is always the bound based on the trivial bound (in the regime in which the lower bound is $>0$). For the upper bound, we consider both the error probability derived from the exact form of the improved lower bound on the trace norm (from Eq.~(\ref{eq: trace norm lower bound})) and the QCRB bound. Since the two bounds overlap almost perfectly, the approximation is valid in this regime ($N=150$). It is clear that there is room for improvement of either the upper or the lower bound on the trace norm for large $N$.

Fig.~\ref{fig: bacteria discrimination} bounds the error probability over time for discriminating between samples of \textit{E. coli} and \textit{Salmonella}. In this scenario, $\mathcal{C}_X$ is the channel corresponding to a sample containing \textit{E. coli} and $\mathcal{C}_Y$ is the channel corresponding to a sample containing \textit{Salmonella}. In this case, we calculate the time-dependent transmissivities differently, by modelling the absorbances, $A$, of the samples as Gompertz functions and applying the formula
\begin{align}
\eta=10^{-A}.
\end{align}
The absorbances are modelled as following
\begin{align}
&A=c_{1}e^{a}+A_{bk},\\
&a=-e^{\frac{c_2 e}{c_1}(c_3-t)+1},
\end{align}
where $A_{\mathrm{bk}}$ is the absorbance of a blank sample and $c_1$, $c_2$, and $c_3$ are experimentally determined coefficients that depend on the type of bacteria present in the sample. Ref.~\cite{spedalieri_detecting_2020} found that the triple $(c_1,c_2,c_3)$ took values $(0.309,0.139,2.634)$ for \textit{E. coli} and $(0.242,0.0882,2.672)$ for \textit{Salmonella}. $A_{bk}$ (which was the same for both samples) took the value $0.144$.

The lower bound is derived from the tightest of our upper bounds on the trace norm, which is again the trivial bound over the entire regime in which the lower bound is $>0$. The upper bounds are calculated in the same way as for Fig.~\ref{fig: bacteria detection}, and we again find that the exact form of the bound and the approximation overlap almost perfectly. The bounds briefly peak after a little more than $2$ hours, before decreasing again, due to the fact that the difference in the absorbances of the samples briefly decreases before increasing again. Once again, we have a large gap between the bounds, which could be improved by tightening either the lower or the upper bound. It is not yet known which bound most needs to be tightened.

\hfill
\section{Conclusion}\label{section: conclusion}
In this paper, we have presented multiple new bounds on the optimal trace norm for discriminating between two AD channels. We have strengthened both the upper and the lower bounds on the optimal trace norm by presenting the improved Choi bounds, the alternative resource bounds, the trivial bound and the improved lower bound on the trace norm. We have also calculated the exact diamond norm between AD channels, thus obtaining the exact error probability for one-shot channel discrimination between any two AD channels, in analytical form.

The bounds were then numerically investigated, and we found that the either the alternative resource bound or the trivial bound gave the tightest lower bound over a wide range of parameter ($\eta_X$, $\eta_Y$ and $N$) values. The bounds were applied to two different scenarios: quantum hacking of BB84 and biological quantum metrology (detecting and discriminating between bacteria in a sample). In the latter scenario, we also confirm that the QCRB bound is valid as an approximation of the discrimination error probability derived from the improved lower bound on the trace norm (and is therefore a valid upper bound on the error probability) for large $N$ (in our case, $N=150$).

We briefly discussed how these results could be extended to pure loss, qudit channels, however this is an area that is open to more research, which could find bounds on the error probability of adaptive discrimination protocols between any two lossy channels. Another area for continued research is the further tightening of either the upper or the lower bound on the optimal trace norm, since there is still room for improvement.

In conclusion, our work contributes to the theory of channel simulation of AD channels and significantly improves the bounds on the optimal error probabilities for adaptive discrimination protocols between AD channels.

\smallskip
\textbf{Acknowledgments.}~This work was funded by the European Union's Horizon 2020 Research and Innovation Action under grant agreement No. 862644 (FET-OPEN project: Quantum readout techniques and technologies, QUARTET) and by the EPSRC via the Quantum Communications Hub (Grant Nos. EP/M013472/1 and EP/T001011/1).


\begin{thebibliography}{43}

\bibitem{pirandola_advances_2018}
S. Pirandola, B.~R. Bardhan, T. Gehring, C. Weedbrook, and S. Lloyd,
\newblock \textit{Advances in photonic quantum sensing},
\newblock Nat. Photon. \textbf{12}, 724 (2018).
  
\bibitem{scarani_black_2014}
V. Scarani, and C. Kurtsiefer,
\newblock \textit{The black paper of quantum cryptography: Real implementation problems},
\newblock Theor. Comput. Sci. \textbf{560}, 27 (2014).

\bibitem{jain_attacks_2016}
N. Jain, B. Stiller, I. Khan, D. Elser, C. Marquardt, and G. Leuchs,
\newblock \textit{Attacks on practical quantum key distribution systems (and how to prevent them)},
\newblock Contemp. Phys. \textbf{57}, 366 (2016).

\bibitem{pirandola_advances_2020}
S. Pirandola, U.~L. Andersen, L. Banchi, M. Berta, D. Bunandar, R. Colbeck, D. Englund, T. Gehring, C. Lupo, C. Ottaviani, J. Pereira, M. Razavi, J.~S. Shaari, M. Tomamichel, V.~C. Usenko, G. Vallone, P. Villoresi, and P. Wallden,
\newblock \textit{Advances in Quantum Cryptography},
\newblock Adv. Opt. Photonics \textbf{12}, 1012 (2020).

\bibitem{helstrom_quantum_1969}
C.~W. Helstrom,
\newblock \textit{Quantum detection and estimation theory},
\newblock J. Stat. Phys. \textbf{1}, 231 (1969).

\bibitem{lloyd_enhanced_2008}
S. Lloyd,
\newblock \textit{Enhanced Sensitivity of Photodetection via Quantum Illumination},
\newblock Science \textbf{321}, 1463 (2008).

\bibitem{tan_quantum_2008}
S.-H. Tan, B.~I. Erkmen, V. Giovannetti, S. Guha, S. Lloyd, L. Maccone, S. Pirandola, and J.~H. Shapiro,
\newblock \textit{Quantum Illumination with Gaussian States},
\newblock Phys. Rev. Lett. \textbf{101}, 253601 (2008).

\bibitem{shapiro_quantum_2009}
J.~H Shapiro and S. Lloyd,
\newblock \textit{Quantum illumination versus coherent-state target detection},
\newblock New J. Phys. \textbf{11}, 063045 (2009).

\bibitem{barzanjeh_microwave_2015}
S. Barzanjeh, S. Guha, C. Weedbrook, D. Vitali, J.~H. Shapiro, and S. Pirandola,
\newblock \textit{Microwave Quantum Illumination},
\newblock Phys. Rev. Lett. \textbf{114}, 080503 (2015).

\bibitem{Zhuang17}
Q. Zhuang, Z. Zhang, and J.~H. Shapiro,
\newblock \textit{Entanglement-enhanced Neyman-Pearson target detection using quantum illumination},
\newblock J. Opt. Soc. Am. B \textbf{34}, 1567 (2017).

\bibitem{Zhuang17b}
Q. Zhuang, Z. Zhang, and J.~H. Shapiro,
\newblock \textit{Quantum illumination for enhanced detection of Rayleigh-fading targets},
\newblock Phys. Rev. A \textbf{96}, 020302(R) (2017).

\bibitem{Wilde17}
M.~M. Wilde, M. Tomamichel, S. Lloyd, and M. Berta,
\newblock \textit{Gaussian hypothesis testing and quantum illumination},
\newblock Phys. Rev. Lett. \textbf{119}, 120501 (2017).

\bibitem{QuntaoLIDARS}
Q. Zhuang, Z. Zhang, and J.~H. Shapiro,
\newblock \textit{Entanglement-enhanced lidars for simultaneous range and velocity measurements},
\newblock Phys. Rev. A \textbf{96}, 040304(R) (2017).

\bibitem{Depalma18}
G. De~Palma and J. Borregaard,
\newblock \textit{Minimum error probability of quantum illumination},
\newblock Phys. Rev. A \textbf{98}, 012101 (2018).

\bibitem{NairGu}
R. Nair and M. Gu,
\newblock \textit{Fundamental limits of quantum illumination},
\newblock Optica \textbf{7}, 771 (2020).

\bibitem{Athena2020}
A. Karsa, G. Spedalieri, Q. Zhuang, and S. Pirandola,
\newblock \textit{Quantum illumination with a generic Gaussian source},
\newblock Phys. Rev. Res. \textbf{2}, 023414 (2020).

\bibitem{Lopaeva13}
E.~D. Lopaeva, I. Ruo~Berchera, I.~P. Degiovanni, S. Olivares, G. Brida, and M. Genovese,
\newblock \textit{Experimental realization of quantum illumination},
\newblock Phys. Rev. Lett. \textbf{110}, 153603 (2013).

\bibitem{ZhangQI13}
Z. Zhang, M. Tengner, T. Zhong, F.~N.~C. Wong, and J.~H. Shapiro,
\newblock \textit{Entanglement's benefit survives an entanglement-breaking channel},
\newblock Phys. Rev. Lett. \textbf{111}, 010501 (2013).

\bibitem{ZhangQI15}
Z. Zhang, S. Mouradian, F.~N.~C. Wong, and J.~H. Shapiro,
\newblock \textit{Entanglement-enhanced sensing in a lossy and noisy environment},
\newblock Phys. Rev. Lett. \textbf{114}, 110506 (2015).

\bibitem{qreading}
S. Pirandola,
\newblock \textit{Quantum reading of a classical digital memory},
\newblock Phys. Rev. Lett. \textbf{106}, 090504 (2011).

\bibitem{harrow_adaptive_2010}
A.~W. Harrow, A. Hassidim, D.~W. Leung, and J. Watrous,
\newblock \textit{Adaptive versus nonadaptive strategies for quantum channel discrimination},
\newblock Phys. Rev. A \textbf{81}, 032339 (2010).

\bibitem{acin_statistical_2001}
A. Ac\'{i}n,
\newblock \textit{Statistical Distinguishability between Unitary Operations},
\newblock Phys. Rev. Lett. \textbf{87}, 177901 (2001).

\bibitem{hayashi_discrimination_2009}
M. Hayashi,
\newblock \textit{Discrimination of Two Channels by Adaptive Methods and Its Application to Quantum System},
\newblock IEEE Trans. Inf. Theory \textbf{55}, 3807 (2009).

\bibitem{pirandola_fundamental_2017}
S. Pirandola, R. Laurenza, C. Ottaviani, and L. Banchi,
\newblock \textit{Fundamental limits of repeaterless quantum communications},
\newblock Nat. Commun. \textbf{8}, 15043 (2017).

\bibitem{pirandola_fundamental_2019}
S. Pirandola, R. Laurenza, C. Lupo, and J.~L. Pereira,
\newblock \textit{Fundamental limits to quantum channel discrimination},
\newblock npj Quantum Inf. \textbf{5}, 50 (2019).

\bibitem{cope_simulation_2017}
T.~P.~W. Cope, L. Hetzel, L. Banchi, and S. Pirandola,
\newblock \textit{Simulation of non-Pauli channels},
\newblock Phys. Rev. A \textbf{96}, 022323 (2017).

\bibitem{pirandola_conditional_2019}
S. Pirandola, R. Laurenza, and L. Banchi,
\newblock \textit{Conditional channel simulation},
\newblock Ann. Phys. (N. Y.) \textbf{400}, 289 (2019).

\bibitem{convex}
L. Banchi, J. Pereira, S. Lloyd, and S. Pirandola,
\newblock \textit{Convex optimization of programmable quantum computers},
\newblock npj Quantum Inf. \textbf{6}, 42 (2020).

\bibitem{Sam94}
S.~L. Braunstein and C.~M. Caves,
\newblock \textit{Statistical distance and the geometry of quantum states},
\newblock Phys. Rev. Lett. \textbf{72}, 3439 (1994).

\bibitem{nielsen_quantum_2011}
M.~A. Nielsen and I.~L. Chuang,
\newblock \textit{Quantum Computation and Quantum Information: 10th Anniversary Edition},
\newblock 10th ed.
\newblock (Cambridge University Press, USA, 2011).

\bibitem{bose_quantum_2003}
S. Bose,
\newblock \textit{Quantum Communication through an Unmodulated Spin Chain},
\newblock  Phys. Rev. Lett. \textbf{91}, 207901 (2003).

\bibitem{zhuang_ultimate_2020}
Q. Zhuang and S. Pirandola,
\newblock \textit{Ultimate Limits for Multiple Quantum Channel Discrimination},
\newblock  Phys. Rev. Lett. \textbf{125}, 080505 (2020).

\bibitem{katariya_evaluating_2020}
V. Katariya and M. Wilde,
\newblock \textit{Evaluating the Advantage of Adaptive Strategies for Quantum Channel Distinguishability},
\newblock arXiv:2001.05376 (2020).

\bibitem{rexiti_discriminating_2021}
M. Rexiti and S. Mancini,
\newblock \textit{Discriminating qubit amplitude damping channels},
\newblock J. Phys. A Math. Theor. \textbf{54}, 165303 (2021).

\bibitem{spedalieri_detecting_2020}
G. Spedalieri, L. Piersimoni, O. Laurino, S.~L. Braunstein, and S. Pirandola,
\newblock \textit{Detecting and tracking bacteria with quantum light},
\newblock Phys. Rev. Res. \textbf{2}, 043260 (2020).

\bibitem{audenaert_discriminating_2007}
K.~M.~R. Audenaert, J. Calsamiglia, R. Mu\~{n}oz-Tapia, E. Bagan, Ll. Masanes, A. Acin, and F. Verstraete,
\newblock \textit{Discriminating States: The Quantum Chernoff Bound},
\newblock Phys. Rev. Lett. \textbf{98}, 160501 (2007).

\bibitem{calsamiglia_quantum_2008}
J. Calsamiglia, R. Mu\~{n}oz-Tapia, Ll. Masanes, A. Acin, and E. Bagan,
\newblock \textit{Quantum Chernoff bound as a measure of distinguishability between density matrices: Application to qubit and Gaussian states},
\newblock Phys. Rev. A \textbf{77}, 032311 (2008).

\bibitem{bennett_teleporting_1993}
C.~H. Bennett, G. Brassard, C. Cr\'{e}peau, R. Jozsa, A. Peres, and W.~K. Wootters,
\newblock \textit{Teleporting an unknown quantum state via dual classical and Einstein-Podolsky-Rosen channels},
\newblock Phys. Rev. Lett. \textbf{70}, 1895 (1993).

\bibitem{bowen_teleportation_2001}
G. Bowen and S. Bose,
\newblock \textit{Teleportation as a Depolarizing Quantum Channel, Relative Entropy, and Classical Capacity},
\newblock Phys. Rev. Lett. \textbf{87}, 267901 (2001).

\bibitem{ishizaka_asymptotic_2008}
S. Ishizaka and T. Hiroshima,
\newblock \textit{Asymptotic Teleportation Scheme as a Universal Programmable Quantum Processor},
\newblock Phys. Rev. Lett. \textbf{101}, 240501 (2008).

\bibitem{ishizaka_quantum_2009}
S. Ishizaka and T. Hiroshima,
\newblock \textit{Quantum teleportation scheme by selecting one of multiple output ports},
\newblock Phys. Rev. A \textbf{79}, 042306 (2009).

\bibitem{pereira_characterising_2021}
J. Pereira, L. Banchi, and S. Pirandola,
\newblock \textit{Characterising port-based teleportation as a universal simulator of qubit channels},
\newblock J. Phys. A Math. Theor. \textbf{54}, 205301 (2021).

\bibitem{watrous_simpler_2013}
J. Watrous,
\newblock \textit{Simpler semidefinite programs for completely bounded norms},
\newblock Chicago Journal of Theoretical Computer Science \textbf{2013} (2013).

\bibitem{adesso_optimal_2009}
G. Adesso, F. Dell'Anno, S. De~Siena, F. Illuminati, and L.~A.~M. Souza,
\newblock \textit{Optimal estimation of losses at the ultimate quantum limit with non-Gaussian states},
\newblock Phys. Rev. A \textbf{79}, 040305(R) (2009).

\bibitem{braunstein_maximum-likelihood_1992}
S.~L. Braunstein, A.~S. Lane, and C.~M. Caves,
\newblock \textit{Maximum-likelihood analysis of multiple quantum phase measurements},
\newblock Phys. Rev. Lett. \textbf{69}, 2153 (1992).

\bibitem{nechita_almost_2018}
I. Nechita, Z. Pucha\l{}a, \L{}. Pawela, and K. \.{Z}yczkowski,
\newblock \textit{Almost all quantum channels are equidistant},
\newblock J. Math. Phys. \textbf{59}, 052201 (2018).

\bibitem{kim_implementation_2008}
Y.~S. Kim, Y.~C Jeong, and Y.~H. Kim,
\newblock \textit{Implementation of polarization-coded free-space BB84 quantum key distribution},
\newblock Laser Phys. \textbf{18}, 810 (2008).

\bibitem{lucamarini_practical_2015}
M. Lucamarini, I. Choi, M.~B. Ward, J.~F. Dynes, Z.~L. Yuan, and A.~J. Shields,
\newblock \textit{Practical Security Bounds Against the Trojan-Horse Attack in Quantum Key Distribution},
\newblock  Phys. Rev. X \textbf{5}, 031030 (2015).

\bibitem{Shirokov}
M.~E. Shirokov,
\newblock \textit{On the energy-constrained diamond norm and its application in quantum information theory},
\newblock Probl. Inform. Transmiss. \textbf{54}, 20 (2018).

\bibitem{winter_energy-constrained_2017}
A. Winter,
\newblock \textit{Energy-constrained diamond norm with applications to the uniform continuity of continuous variable channel capacities},
\newblock arXiv:1712.10267 (2017).

\bibitem{DATA} Supplementary data, including the code used for the various plots, is available at doi.org/10.5281/zenodo.4281118.
\end{thebibliography}
\end{document}